\documentclass{aa}

\pdfoutput=1
\usepackage{savesym}
\usepackage{amsmath}
\savesymbol{iint}
\usepackage[varg]{txfonts}
\restoresymbol{TXF}{iint}
\usepackage{longtable}
\usepackage{multicol}
\usepackage{wrapfig}
\usepackage{adjustbox}
\usepackage[colorinlistoftodos]{todonotes}
\usepackage{natbib,twoopt}
\bibpunct{(}{)}{;}{a}{}{,} 
\usepackage{hyperref} 
\bibpunct{(}{)}{;}{a}{}{,}             
\makeatletter
  \newcommandtwoopt{\citeads}[3][][]{\href{http://adsabs.harvard.edu/abs/#3}%
    {\def\hyper@linkstart##1##2{}%
     \let\hyper@linkend\@empty\citealp[#1][#2]{#3}}}
  \newcommandtwoopt{\citepads}[3][][]{\href{http://adsabs.harvard.edu/abs/#3}%
    {\def\hyper@linkstart##1##2{}%
     \let\hyper@linkend\@empty\citep[#1][#2]{#3}}}
  \newcommandtwoopt{\citetads}[3][][]{\href{http://adsabs.harvard.edu/abs/#3}%
    {\def\hyper@linkstart##1##2{}%
     \let\hyper@linkend\@empty\citet[#1][#2]{#3}}}
  \newcommandtwoopt{\citeyearads}[3][][]%
    {\href{http://adsabs.harvard.edu/abs/#3}
    {\def\hyper@linkstart##1##2{}%
     \let\hyper@linkend\@empty\citeyear[#1][#2]{#3}}}
\makeatother

\usepackage{tikz}
\usetikzlibrary{shapes, arrows, positioning}

\definecolor{dgreen}{rgb}{0, 0.7, 0}
\definecolor{lgray}{rgb}{0.8, 0.8, 0.8}

\begin{document}

\title{The ALMA-PILS survey: The first detection of doubly-deuterated methyl formate (CHD$_2$OCHO) in the ISM}

\institute{Centre for Star and Planet Formation, Niels Bohr Institute \& Natural History Museum of Denmark, University of Copenhagen, \O ster Voldgade 5–7, DK-1350 Copenhagen K., Denmark\label{inst-NBI} \leavevmode\\ \email{sebastien@nbi.ku.dk}\and INAF, Osservatorio Astrofisico di Arcetri, Largo E. Fermi 5, 50125 Firenze, Italy \label{inst-INAF} \and I. Physikalisches Institut, Universit{\"a}t zu K{\"o}ln, Z{\"u}lpicher Str. 77, 50937 K{\"o}ln, Germany\label{inst-Koln} \and Laboratoire d'Astrophysique de Bordeaux, Univ. Bordeaux, CNRS, B18N, all{\'e} Geoffroy Saint-Hilaire, 33615 Pessac, France \label{inst-LAB} \and Center for Space and Habitability (CSH), University of Bern, Sidlestrasse 5, 3012 Bern, Switzerland \label{inst-Bern} \and ASTRON Netherlands Institute for Radio Astronomy, PO Box 2, 7990 AA Dwingeloo, The Netherlands \label{inst-ASTRON} \and SKA Organisation, Jodrell Bank Observatory, Lower Withington, Macclesfield, Cheshire SK11 9DL, UK\label{inst-SKA}}

\author{S. Manigand$^{\ref{inst-NBI}}$\and H. Calcutt$^{\ref{inst-NBI}}$\and J. K. J\o rgensen$^{\ref{inst-NBI}}$\and V. Taquet$^{\ref{inst-INAF}}$ \and H. S. P. M{\"u}ller$^{\ref{inst-Koln}}$ \and A. Coutens$^{\ref{inst-LAB}}$ \and S. F. Wampfler$^{\ref{inst-Bern}}$ \and N. F. W. Ligterink$^{\ref{inst-Bern}}$ \and M. N. Drozdovskaya$^{\ref{inst-Bern}}$ \and L. E. Kristensen$^{\ref{inst-NBI}}$ \and M. H. D. van der Wiel$^{\ref{inst-ASTRON}}$ \and T. L. Bourke$^{\ref{inst-SKA}}$}

\date{Received xxxx / Accepted 20 November 2018}

\abstract
{Studies of deuterated isotopologues of complex organic molecules can provide important constraints on their origin in regions of star formation. In particular, the abundances of deuterated species are very sensitive to the physical conditions in the environment where they form. Due to the low temperatures in regions of star formation, these isotopologues are enhanced to significant levels, making detections of multiply-deuterated species possible. However, for complex organic species, only the multiply-deuterated variants of methanol and methyl cyanide have been reported so far. The aim of this paper is to initiate the characterisation of multiply-deuterated variants of complex organic species with the first detection of doubly-deuterated methyl formate, CHD$_2$OCHO. We use ALMA observations from the Protostellar Interferometric Line Survey (PILS) of the protostellar binary IRAS~16293--2422, in the spectral range of 329.1 GHz to 362.9 GHz. Spectra towards each of the two protostars are extracted and analysed using a LTE model in order to derive the abundances of methyl formate and its deuterated variants. We report the first detection of doubly-deuterated methyl formate CHD$_2$OCHO in the ISM. The D/H ratio of CHD$_2$OCHO is found to be 2--3 times higher than the D/H ratio of CH$_2$DOCHO for both sources, similar to the results for formaldehyde from the same dataset. The observations are compared to a gas-grain chemical network coupled to a dynamical physical model, tracing the evolution of a molecular cloud until the end of the Class 0 protostellar stage. The overall D/H ratio enhancements found in the observations are of the same order of magnitude as the predictions from the model for the early stages of Class 0 protostars. However, the higher D/H ratio of CHD$_2$OCHO compared to the D/H ratio of CH$_2$DOCHO is still not predicted by the model. This suggests that a mechanism is enhancing the D/H ratio of singly- and doubly-deuterated methyl formate that is not in the model, e.g. mechanisms for H--D substitutions. This new detection provides an important constraint on the formation routes of methyl formate and outlines a path forward in terms of using these ratios to determine the formation of organic molecules through observations of differently deuterated isotopologues towards embedded protostars.}


\titlerunning{The first detection of doubly-deuterated methyl formate (CHD$_2$OCHO) in the ISM}
\authorrunning{S. Manigand et al.}

\maketitle

\section{Introduction}

The earliest stages of star formation are characterised by a rich and abundant molecular diversity in various environments. Complex organic molecules \citep[COMs; molecules with 6 atoms or more containing at least one carbon and hydrogen,][]{Herbst-2009} are particularly abundant in the gas phase close to the protostars, typically at $R < 100$ AU, where the dust temperature is high enough for the ice to sublimate. These inner warm regions around low-mass protostars, known as hot corinos \citep{Ceccarelli-2004}, also show enhanced D/H ratios, or deuterium fractionation, with respect to the local galactic value of $2.0\pm0.1\times10^{-5}$ \citep[][]{Prodanovic-2010}. These ratios are expected to be sensitive to the physical properties and evolutionary stage of the protostar and the chemical formation of specific species \citep{Ceccarelli-2014-conf, Punanova-2016}. Therefore, inventories of their isotopic content may shed further light on the formation routes and the formation time of these important COMs.

The low-mass Class 0 protostellar binary IRAS~16293--2422 (IRAS16293 hereafter) is considered one of the template sources for astrochemical studies. It is known to have a very rich spectrum with bright lines and abundant COMs  \citep[e.g.][]{Blake-1994, Van-Dishoeck-1995, Cazaux-2003, Caux-2011, Jorgensen-2012}. IRAS16293 is located in the nearby Ophiuchus molecular cloud complex at a distance of 141~pc \citep{Dzib-2018} and composed of two protostars A and B separated by 5\farcs1 (720~AU). The two components of the binary show quite different morphologies: the comparatively more luminous south-eastern source (IRAS16293A, hereafter source A), shows an edge-on disk-like morphology, with an elongated dust continuum emission and a velocity gradient of $\sim$6 km~s$^{-1}$. In contrast, the north-western source (IRAS16293B, hereafter source B) shows roughly circular dust continuum emission and narrow lines (FWHM$\approx$1~km~s$^{-1}$), a factor of three narrower than IRAS16293A (i.e. FWHM$\approx$2.5~km~s$^{-1}$) indicating a more face-on morphology.

This study is focused on deuterated COMs and uses data from the Protostellar Interferometric Line Survey \citep[PILS;][]{Jorgensen-2016}, an ALMA survey of IRAS16293. 
\textcolor{black}{Previous single-dish studies of IRAS16293 have shown high D/H ratios for relatively small molecules such as methanol (CH$_3$OH) and formaldehyde (H$_2$CO), although these values have high uncertainties due to unresolved single-dish observations \citep{Loinard-2000, Parise-2002}}. The high angular resolution and sensitivity measurements from PILS opens up a more detailed analysis of the D/H ratio for different species. The PILS data for IRAS16293B have for example been used to demonstrate that the D/H ratios for HDCO and D$_2$CO differ (3.2\% and 9.1\%, respectively; \citealt{Persson-2018}) and also provided measurements of mono-deuterated larger COMs, e.g. molecules formed from CH$_3$O or HCO radicals, such as methyl formate, glycolaldehyde (CH$_2$OHCHO), ethanol (C$_2$H$_5$OH) and dimethyl ether (CH$_3$OCH$_3$), reported with D/H ratios of 4 to 8\% \citep{Jorgensen-2018} and formamide (NH$_2$CHO), with a D/H ratio of $\approx$2\% \citep{Coutens-2016}. The high sensitivity of the dataset enables a search to be undertaken for multiply-deuterated variants of these more complex COMs and investigate whether similar differences are present in the deuteration levels between the mono- and multiply-deuterated variants as an imprint of their formation. 

In this paper, we report the first detection of a doubly-deuterated isotopologue of methyl formate (CHD$_2$OCHO) in the interstellar medium (ISM), towards both components of the IRAS16293 protostellar binary. In addition, the other methyl formate deuterated isotopologues are identified towards the two protostars of the binary system, and we compare the derived D/H ratios with predictions based on ice grain chemistry.

\section{Observations \& spectroscopic data}

The ALMA observations cover a frequency range of 329.1 GHz to 362.9 GHz, with a spectral resolution of 0.2 km\,s$^{-1}$, or 0.244 MHz, and a restoring beam of 0\farcs 5. The data reach a sensitivity of \textcolor{black}{$\sim$5--7 mJy\,beam$^{-1}$\,km\,s$^{-1}$} across the entire frequency range, increasing the sensitivity by one to two orders of magnitude compared to previous surveys. The phase centre is located between the two continuum sources at $\alpha_\mathrm{J2000}$=$16^\mathrm{h} 32^\mathrm{m} 22\fs72$ and $\delta_\mathrm{J2000}$=$-24^\circ 28' 34\farcs3$. Further details of the data reduction and continuum subtraction can be found in  \cite{Jorgensen-2016}. Two positions, offset by 0{\farcs}6 and 0{\farcs}5 from the peak continuum positions, have been investigated towards IRAS16293A and IRAS16293B, respectively. The offset positions are the same as those used in previous PILS studies \citep{Lykke-2017, Ligterink-2017, Calcutt-2018a}. The spectroscopic data used in this study are given in Appendix \ref{App-spectro}.

\section{\label{sec_results}Results \& Analysis}

\begin{figure}[t]
\begin{tabular}{p{0.04\textwidth} p{0.21\textwidth} p{0.23\textwidth}}
& {IRAS16293A} & {IRAS16293B}
\end{tabular}

\adjustbox{trim ={0.02\width} {0.05\height} {0.02\width} 0, clip=true}{\includegraphics[scale=0.64]{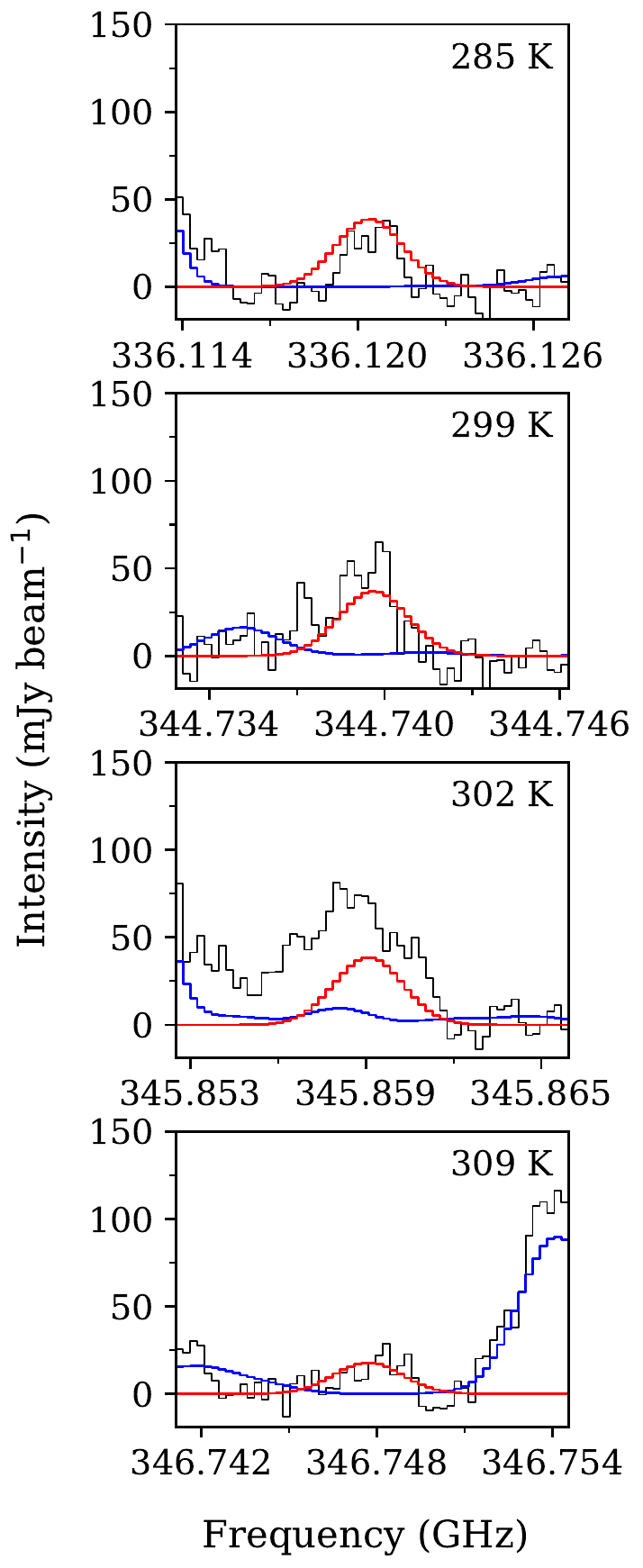}}
\hspace{5pt}
\adjustbox{trim ={0.1\width} {0.05\height} {0.02\width} 0, clip=true}{\includegraphics[scale=0.64]{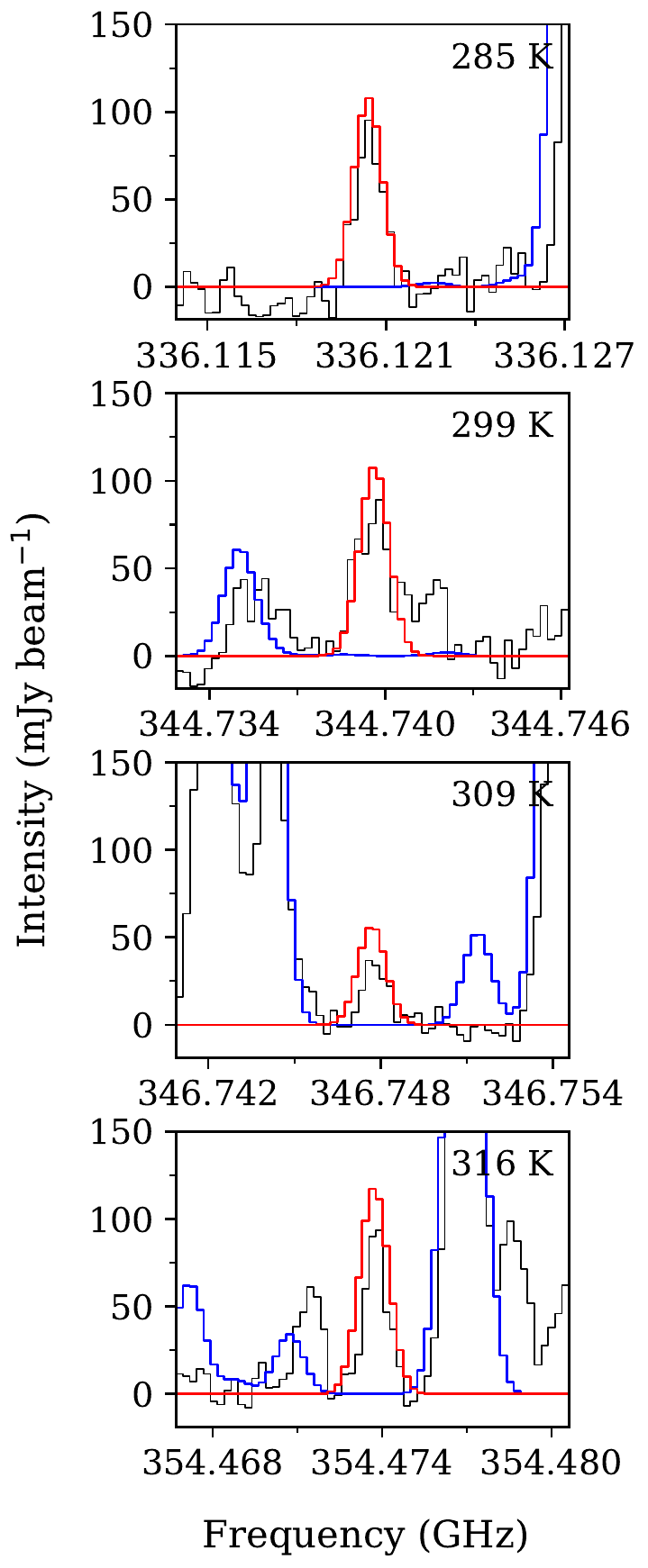}}\\
{\textcolor{white}{oooooooooooooooooo}\small Frequency (GHz)}
\caption{\label{fig-synth}Selection of the brightest CHD$_2$OCHO transitions towards IRAS16293A (left) and IRAS16293B (right). The synthetic spectra are over-plotted in red, a reference spectrum in blue and the data in black. The upper-level energies are indicated in the top-right corner of each plot.}
\end{figure}

We searched for CHD$_2$OCHO towards both sources, and for CH$_2$DOCHO, CH$_3$OCDO and CH$_3$OCHO towards source A; \cite{Jorgensen-2018} reported on the last three isotopologues towards source B.
To identify the presence of the species in the gas, a Gaussian line profile synthetic spectrum was used and compared to extracted spectra towards both sources of the binary system. Given the high density of the gas at the scales of the PILS data, the gas of the inner region is assumed to be in Local Thermodynamic Equilibrium \citep[LTE, see][]{Jorgensen-2016}. 

\indent 54 lines of CHD$_2$OCHO were identified above 3$\sigma$, including 18 unblended with other species, towards source B, and 5 lines above 3$\sigma$, including \textcolor{black}{2 unblended lines}, towards source A. In addition, 419, 82 and 3 unblended lines above the 3$\sigma$ detection threshold were identified for CH$_3$OCHO, CH$_2$DOCHO and CH$_3$OCDO, respectively, towards IRAS16293A. A significant number of lines of the main isotopologue (i.e. 260 lines) were found to be optically thick and excluded in the subsequent fits.

\indent A large $\chi^2$ grid of spectral models was run in order to constrain 
the excitation temperature, $T_\mathrm{ex}$, the column density, $N_\mathrm{tot}$, the peak velocity, $V_\mathrm{peak}$, and the line full width at half maximum, FWHM, that best reproduce the extracted spectra. The grid covered column density values from $1\times10^{15}$ to $1\times10^{18}$ cm$^{-2}$
and excitation temperatures from 100 to 300~K. Ranges of 2.0 to 2.6 km s$^{-1}$ and 0.6 to 1.0 km s$^{-1}$ for FWHM and $V_\mathrm{peak}$ respectively, were used for source A. The ranges for source B were 0.8 to 1.2 km s$^{-1}$ and 2.5 to 2.9 km s$^{-1}$ for FWHM and $V_\mathrm{peak}$, respectively. For the A source, the FWHM and the peak velocity are $2.4\pm0.2$ km\,s$^{-1}$ and $0.8\pm0.2$ km\,s$^{-1}$, respectively, which is in good agreement with the N-bearing molecules \citep[FWHM = 2.2 and 2.5 km\,s$^{-1}$ and $V_\mathrm{peak}$ = 0.8 and 0.8 km\,s$^{-1}$ from][respectively]{Calcutt-2018a, Ligterink-2017}. 
The FWHM and the peak velocity are found to be $1.0\pm0.2$ km\,s$^{-1}$ and $2.7\pm0.2$ km\,s$^{-1}$, respectively, for the B source, similar to previous studies on the same dataset \citep[e.g.][]{Coutens-2016, Lykke-2017, Drozdovskaya-2018}. 
Each line is weighted with the spectral resolution and a factor corresponding to the risk of blending effect with other species.
In all models, the same source size $\theta_\mathrm{s}$ =  0\farcs 5 was adopted. This value corresponds to the deconvolved extent of the marginally resolved sources in the beam of these ALMA observations \citep{Jorgensen-2016}. This source size was also adopted in all previous studies using the PILS dataset. A reference spectrum based on the column density of the detected species in previous publications on the same dataset was used to reduce blending effects, due to the larger line widths towards IRAS16293A and the high line concentration in the spectra. 

\begin{table}[t]
\centering
\caption{\label{tab_results}Derived column densities of the methyl formate isotopologues towards offset positions (see Section \ref{sec_results}).}
\begin{tabular}{lcc}
\hline\hline
Species  & \multicolumn{2}{c}{ $N_{\mathrm{tot}}$ (cm$^{-2}$)}\\
 & IRAS16293A & IRAS16293B \\
\hline
\small CHD$_2$OCHO  & $5.3\pm1.6\times10^{15}$  & $1.1\pm0.2\times10^{16}$ \\
\small CH$_2$DOCHO  & $2.3\pm0.7\times10^{16}$ & $4.8\pm0.5\times10^{16\dagger}$  \\
\small CH$_3$OCDO  & $4.5\pm1.3\times10^{15}$ & $1.5\pm0.2\times10^{16\dagger}$  \\
\small CH$_3$OCHO  & $2.7\pm0.8\times10^{17}$ & $2.6\pm0.3\times10^{17\dagger}$ \\
\hline
\end{tabular}
\tablebib{
$\dagger$: \cite{Jorgensen-2018}. }
\end{table}
\indent The best fit for the excitation temperature is $115 \pm 25$~K for CH$_3$OCHO towards the offset position of IRAS16293A. This value is significantly different from the excitation temperature found towards IRAS16293B for the same molecules \citep[i.e. 300~K,][]{Jorgensen-2018}. The low dispersion in the upper state energy values, $E_\mathrm{u}$, of the most intense and unblended CHD$_2$OCHO lines, between 260 and 320~K, means that the excitation temperature cannot be constrained properly. Therefore, we used the same excitation temperature as for the main species towards both sources, respectively. 
Anti-coincidences between the synthetic and observed spectrum were searched, but none were found, suggesting that these excitation temperatures provide a good fit. 
Unless the excitation temperatures are differing between the various isotopologues, their derived relative abundances do not significantly depend on the exact excitation temperature considered. 

Figure \ref{fig-synth} shows the brightest CHD$_2$OCHO lines towards the two sources. In general, the fit is better towards the B source than towards the A source, because of the narrower line widths, which reduce the confusion due to line blending. More fitted lines of CH$_3$OCHO, CH$_2$DOCHO, CH$_3$OCDO and CHD$_2$OCHO are shown in Appendix \ref{App-spec} and \ref{App-d-spec}.

Table \ref{tab_results} shows the derived column densities of CH$_3$OCHO, CH$_2$DOCHO, CH$_3$OCDO and CHD$_2$OCHO towards IRAS16293 A and B.
In order to take into account the continuum contribution to the line emission as a background temperature, a correction factor was determined to correct the derived column density, based on the excitation temperature and the continuum flux \citep[see][]{Calcutt-2018b}. At the 0\farcs6 offset position towards the A source, the continuum flux is 0.55 Jy\,beam$^{-1}$, corresponding to a brightness temperature of 30 K, and a correction factor of 1.27 at 115 K. The continuum correction factor at 300 K for the 0\farcs5 offset position towards the B source was 1.05 \citep{Jorgensen-2016}. 

\begin{figure}[t]
\begin{tikzpicture}[auto, node distance=1cm, scale = 1.0]
\node[left] at (0,0) {\adjustbox{trim ={0.13\width} {0.12\height} 0 0, clip=true}{\includegraphics[scale=0.56]{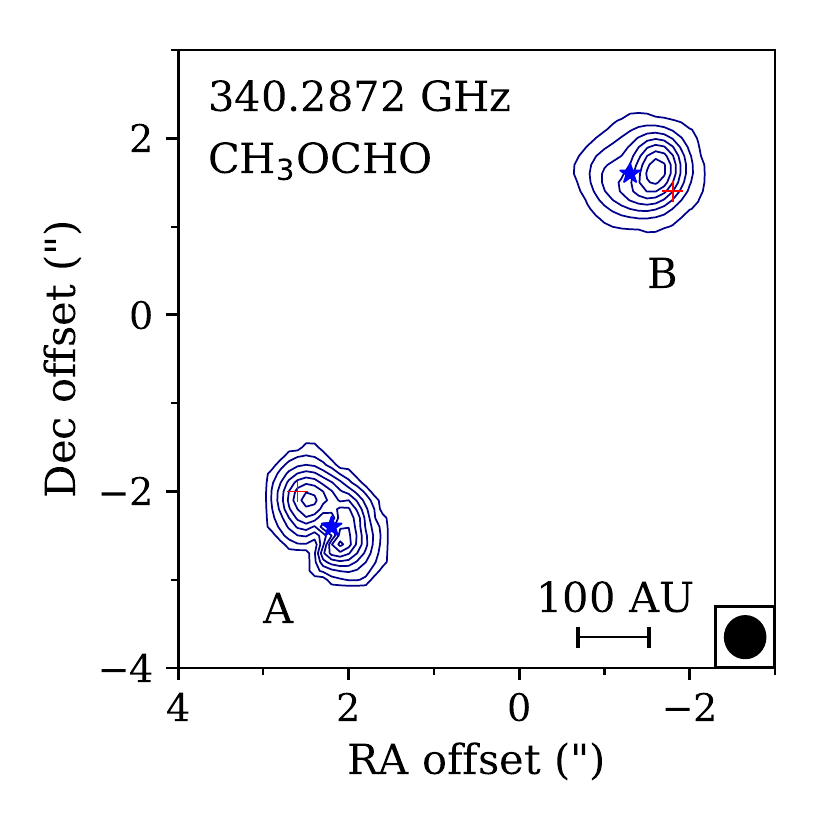}}};
\node[right] at (0,0) {\adjustbox{trim ={0.13\width} {0.12\height} 0 0, clip=true}{\includegraphics[scale=0.56]{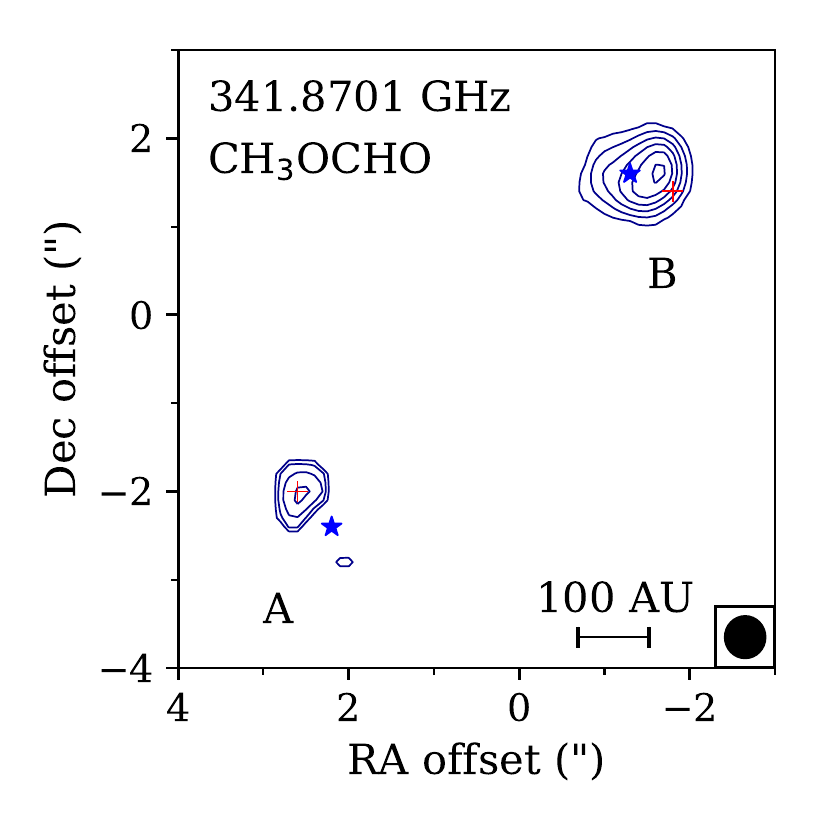}}};
\node[left] at (0,-4.3) {\adjustbox{trim ={0.13\width} {0.12\height} 0 0, clip=true}{\includegraphics[scale=0.56]{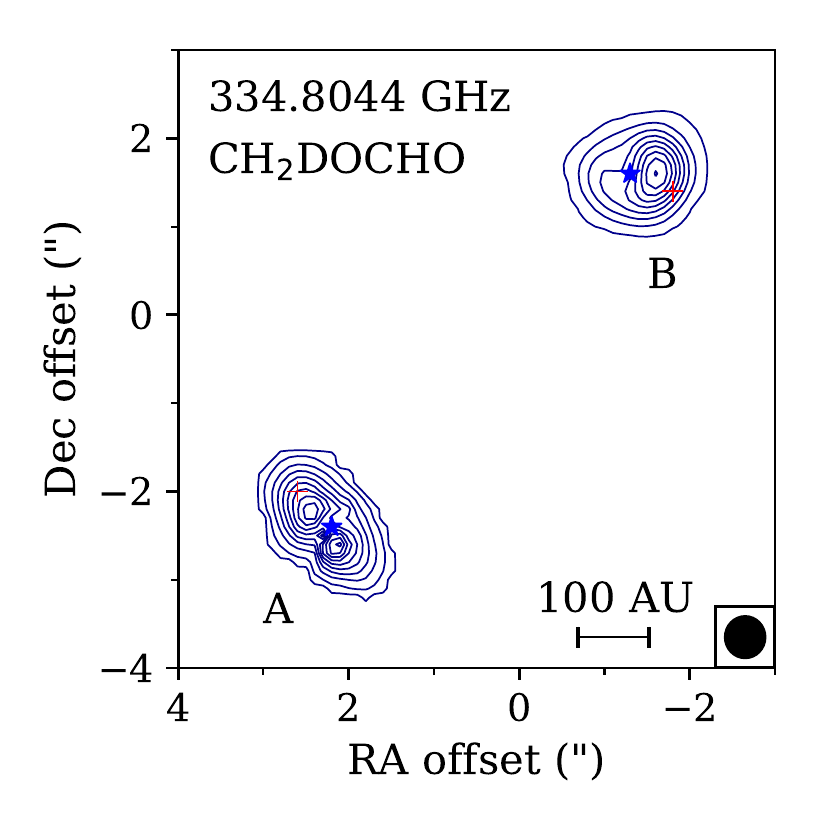}}};
\node[right] at (0,-4.3) {\adjustbox{trim ={0.13\width} {0.12\height} 0 0, clip=true}{\includegraphics[scale=0.56]{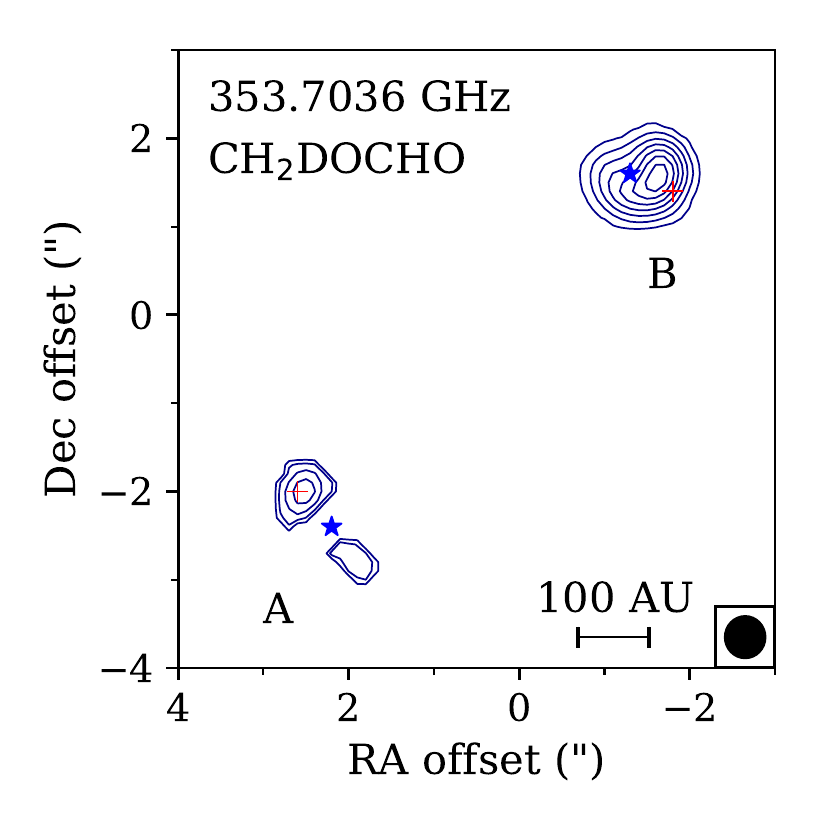}}};
\node[left] at (0,-8.6) {\adjustbox{trim ={0.13\width} {0.12\height} 0 0, clip=true}{\includegraphics[scale=0.56]{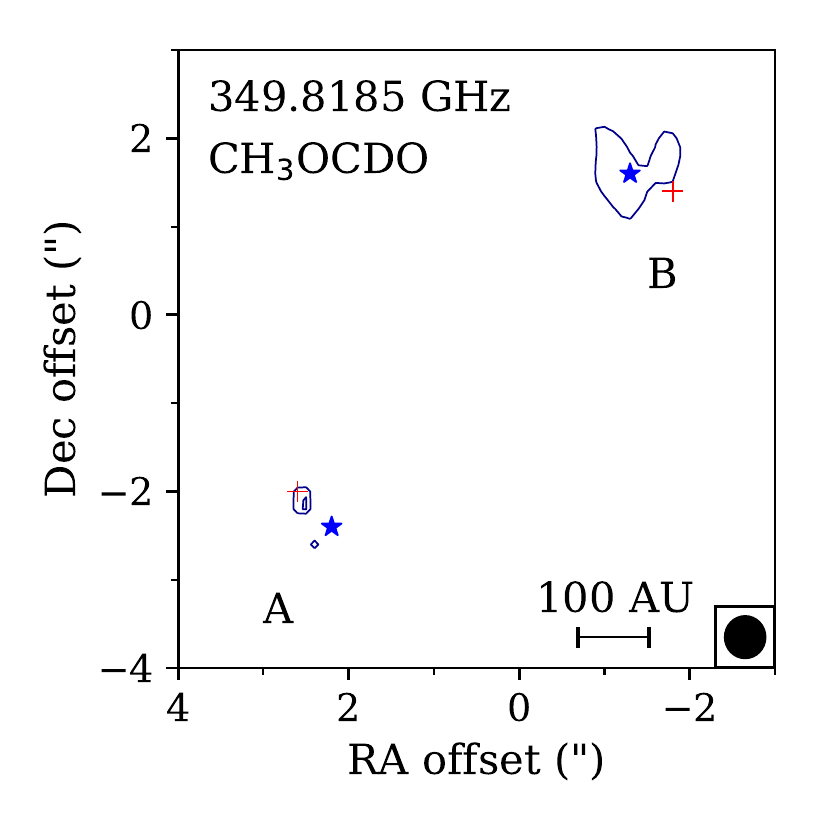}}};
\node[right] at (0,-8.6) {\adjustbox{trim ={0.13\width} {0.12\height} 0 0, clip=true}{\includegraphics[scale=0.56]{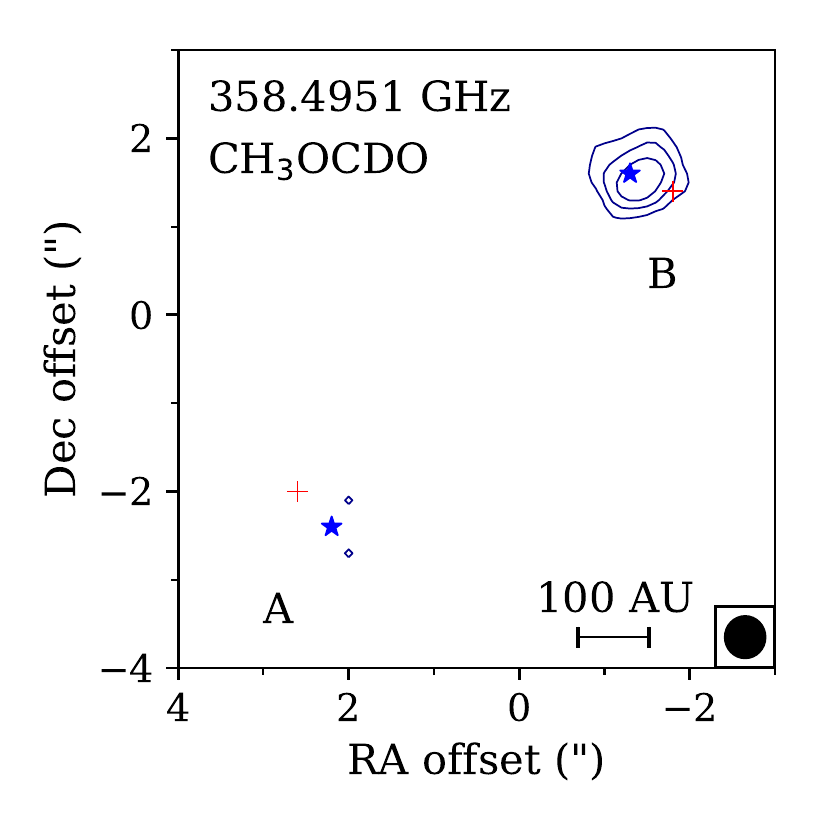}}};
\node[left] at (0,-12.9) {\adjustbox{trim ={0.13\width} {0.12\height} 0 0, clip=true}{\includegraphics[scale=0.56]{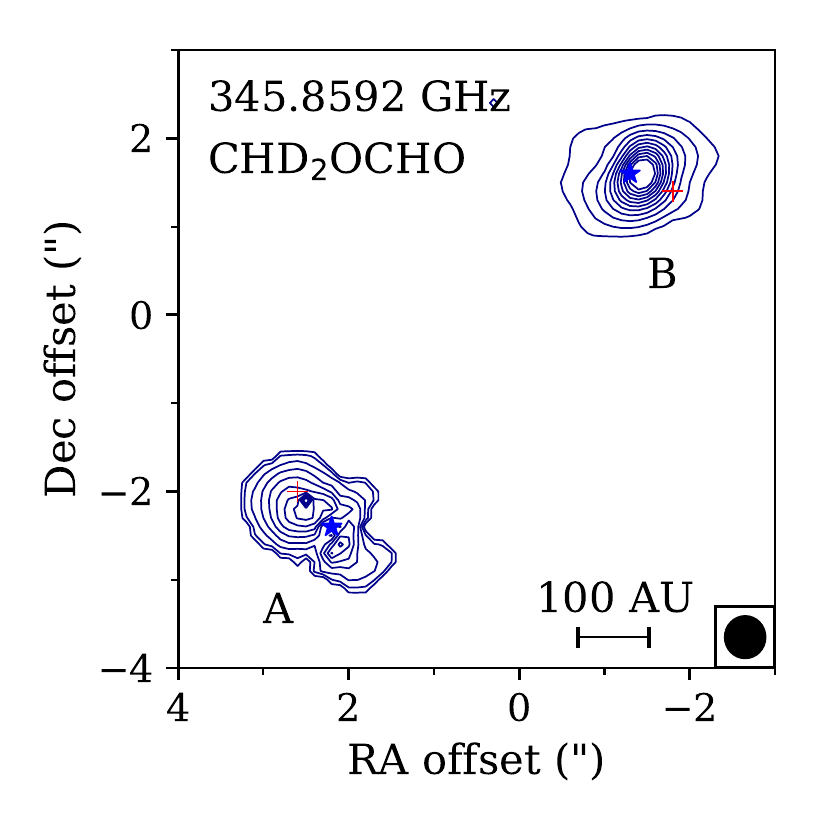}}};
\node[right] at (0,-12.9) {\adjustbox{trim ={0.13\width} {0.12\height} 0 0, clip=true}{\includegraphics[scale=0.56]{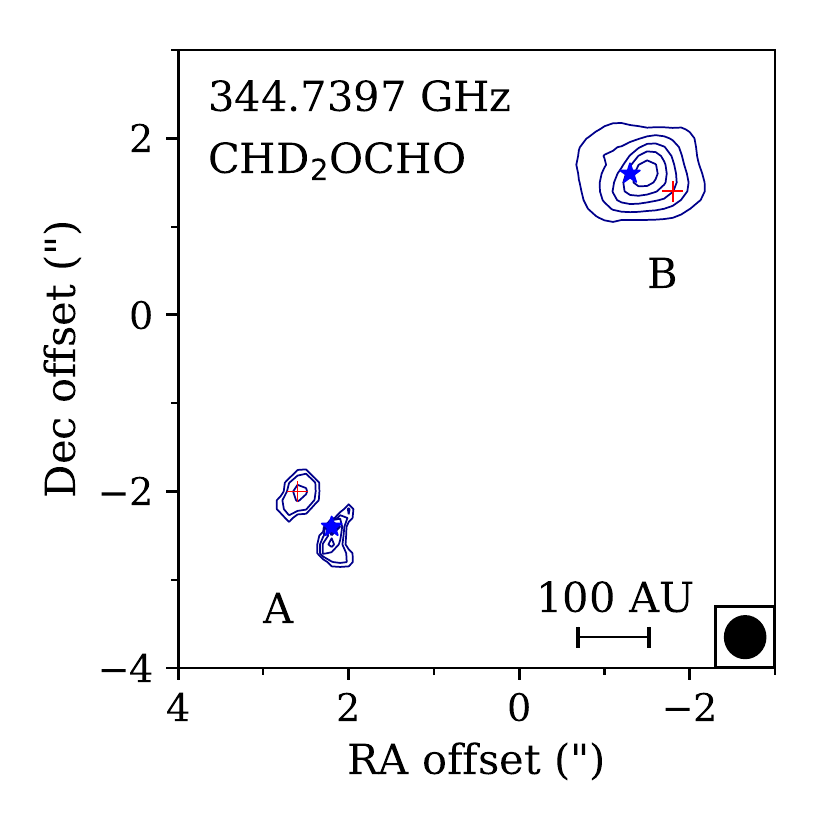}}};
\node[below] at (0,-15.05) {\small RA offset (")};
\node[left] at (-4.3,-7.5) {\rotatebox{90}{\small Dec offset (")}};
\end{tikzpicture}
\caption{\label{fig-vcim} VINE maps of two of the brightest transitions of each methyl formate isotopologue. Contour levels start at 3$\sigma$, with $\sigma$ = 10 mJy\,beam$^{-1}$\,km\,s$^{-1}$, and increase in steps of 3$\sigma$. Continuum peak positions and extracted spectra positions of IRAS16293A and B are indicated with blue stars and red crosses, respectively.}
\end{figure}

\begin{table*}[t]
\caption{\label{tab_abun}Abundance and D/H ratios of methyl formate and formaldehyde isotopologues observed towards IRAS16293A and B and in the model used in the discussion.}
\centering
\begin{tabular}{l cc cc}
\hline\hline
 & \multicolumn{2}{c}{IRAS16293A} & \multicolumn{2}{c}{IRAS16293B}\\
Species & N$_\text{tot}$ ratios ($\times10^{-2}$) & D/H ratios ($\times10^{-2}$) & N$_\text{tot}$ ratios ($\times10^{-2}$) & D/H ratios ($\times10^{-2}$) \\
\hline
CH$_2$DOCHO{\small\ /CH$_3$OCHO} & $8.5 \pm 3.6$ & $2.8 \pm 0.4$ & $18 \pm 2.7\ ^{(1)}$ & $6.1 \pm 0.3\ ^{(1)}$  \\
CH$_3$OCDO{\small\ /CH$_3$OCHO} & $1.7 \pm 0.7$ & $1.7 \pm 0.7$ & $5.7 \pm 0.9\ ^{(1)}$ & $5.7 \pm 0.9\ ^{(1)}$  \\
CHD$_2$OCHO{\small\ /CH$_3$OCHO} & $2.0 \pm 0.8$ & $8.2 \pm 0.6$ & $4.2 \pm 0.6$ & $12.0 \pm 0.3$  \\
HDCO{\small\ /H$_2$CO} & - & - & $6.8 \pm 1.2\ ^{(2)}$ & $3.3 \pm 1.0\ ^{(2)}$  \\
D$_2$CO{\small\ /H$_2$CO} & - & - & $0.9 \pm 0.2\ ^{(2)}$ & $9.2 \pm 1.0\ ^{(2)}$  \\
\hline
 & \multicolumn{2}{c}{model$^{(3)}$ "early time"} & \multicolumn{2}{c}{model$^{(3)}$ "late time"}\\
Species & N$_\text{tot}$ ratios ($\times10^{-2}$) & D/H ratios ($\times10^{-2}$) & N$_\text{tot}$ ratios ($\times10^{-2}$) & D/H ratios ($\times10^{-2}$) \\
\hline
CH$_2$DOCHO{\small\ /CH$_3$OCHO} & 21 & 7.0 & 0.65 & 0.22 \\
CH$_3$OCDO{\small\ /CH$_3$OCHO} & 9.2 & 9.2 & 0.0025 & 0.0025 \\
CHD$_2$OCHO{\small\ /CH$_3$OCHO} & 2.4$^{(*)}$ & 8.9$^{(*)}$ & 0.0016$^{(*)}$ & 0.19$^{(*)}$ \\
HDCO{\small\ /H$_2$CO} & 4.1 & 2.1 & 0.42 & 0.21 \\
D$_2$CO{\small\ /H$_2$CO} & 0.18 & 2.4 & 0.00064 & 0.15 \\
\hline
\end{tabular}
\tablebib{
(1) \cite{Jorgensen-2018}; (2) \cite{Persson-2018}; (3) \cite{Taquet-2014}; $(*)$ V. Taquet, private communication.
}
\end{table*}

\indent \textcolor{black}{The spatial extent of the molecular emission towards IRAS16293A is particularly difficult to explore because of its geometry, in comparison to the B source.}
The 6 km\,s$^{-1}$ velocity gradient in the NE-SW axis \citep{Pineda-2012, Favre-2014-struct}, and the high concentration of lines make standard integrated emission maps almost impossible without contamination by other molecular lines. \cite{Calcutt-2018a} developed a new method to analyse the spatial emission of such velocity structures, called `\textit{Velocity-corrected INtegrated Emission maps}' (VINE maps hereafter). Instead of integrating the same channels over the map, they use the peak velocity at each pixel of a bright methanol line, the $J_K = 7_{3} \rightarrow 6_{3},\ \varv _t = 1$ line of the E symmetrical conformer at 337.519 GHz, as a reference to shift the integration channel range. This methanol transition traces the velocity gradient of the source and hence is suitable to use for the velocity correction \citep[see][ for more details on the method and the velocity map of the reference transition in Figure 5 of their paper]{Calcutt-2018a}.

\indent Figure \ref{fig-vcim} shows the VINE map of two unblended transitions of each deuterated isotopologue of methyl formate. The integration ranges are 3.2 and 1.2 km s$^{-1}$ centered on the corrected peak velocity towards the sources A and B, respectively. Towards IRAS16293B, methyl formate isotopologues spatial distribution traces the most compact region of the protostellar envelope, corresponding to the hot corino region.
Towards source A, the distribution shows a doubly-peaked structure along the velocity gradient NE--SW axis that could correspond to the continuum emission feature at 1~cm previously reported \citep[e.g.][]{Wootten-1989, Chandler-2005, Pech-2010} and shown by the N-bearing species \citep{Calcutt-2018a}. Further VINE maps of other transitions of CH$_3$OCHO and their deuterated isotopologues are given in Appendix \ref{App-vciem}.

\section{Astronomical Implications}

This study presents the first detection of a doubly-deuterated large complex organic molecule, i.e. with a longer main chain than methanol, making it unique.
Table \ref{tab_abun} shows the D/H ratio of CH$_2$DOCHO and CHD$_2$OCHO, determined from the abundance ratios (see Appendix \ref{sec-stat-corr}).
Their D/H ratio is enhanced compared to the local D/H ratio of $2.0\pm0.1\times10^{-5}$ in the ISM \citep{Prodanovic-2010}, as observed for most of the organic molecules towards hot corinos \citep[e.g.][]{Parise-2006}.
The D/H ratio of CHD$_2$OCHO is also enhanced towards both sources, compared to the D/H ratio of CH$_2$DOCHO. The enhanced D/H ratio of the doubly-deuterated isotopologue with respect to the singly deuterated D/H ratio is also seen for formaldehyde towards the B source \citep{Persson-2018}. 

To put these results into context, we compared the abundance to the predictions of the gas-grain chemical model of the earliest stages of hot core formation from \cite{Taquet-2014}. The model couples the astrochemical model GRAINOBLE \citep{Taquet-2012} with a one-dimensional evolutionary model of core collapse. The physical model starts at the molecular cloud stage and then follows the static formation of the dense prestellar core. When the central density reaches $2\times10^{5}$ cm$^{-3}$, the free-fall core collapse is triggered and the time defined as t = 0. After one free-fall timescale ($\sim$9.4$\ \times10^4$ yrs), a protostar is formed and heats up the protostellar envelope, while the envelope shells of increasing radii are gradually accreted at increasing times in an inside-out manner. The end of the Class 0 protostellar stage is defined as the point when the central protostar has accreted half of the total (envelope+protostellar) mass in the system. The kinetic temperature in the deeply embedded envelope during the prestellar phase is lower than 10 K and only increases once the protostar is formed. The temperature profile during this stage is computed using the 1D dust radiative transfer code DUSTY \citep{Ivezic-1997}.  

\begin{figure}[t]
\begin{tikzpicture}[auto, node distance=1cm, scale = 0.8]

	\draw (-4.3,0.0) rectangle (-1.7,-0.6);
	\draw (-1.3,0.0) rectangle (1.3, -0.6);
	\draw (1.7, 0.0) rectangle (4.3, -0.6);	
	\draw (2.2, 1.4) rectangle (3.8, 0.8);	
	\draw (-2.5, 2.6) rectangle (-0.5, 2.0);	
	\draw (0.5, 2.6) rectangle (2.5, 2.0);	
	\draw (3.6, 2.6) rectangle (5.4, 2.0);	
	\draw (4.75, 4.4) rectangle (6.0, 3.8);	
	\draw (-2.2, 5.6) rectangle (-0.8, 5.0);	
	\draw (0.8, 5.6) rectangle (2.2, 5.0);	
	\draw (3.9, 5.6) rectangle (5.1, 5.0);	
	\draw (2.5, 6.2) rectangle (3.5, 5.7);	
	\draw (1.1, 7.0) rectangle (1.9, 6.5);	
	\draw (-3.0,0) node[below] { CHD$_2$OCHO};
	\draw (0,0) node[below] { CH$_2$DOCHO};
	\draw (3.0,0) node[below] { CH$_3$OCHO};
	\draw[->, thick, color=black] (-3.0,0.8) -- (-3.0,0);
	\draw[->, thick, color=black] (0,0.8) -- (0,0);
	\draw[->, thick, color=black] (3.0,0.8) -- (3.0,0);
	\draw[color=black] (-3.0,0.4) node[right] {\small + HCO};
	\draw[color=black] (0,0.4) node[right] {\small + HCO};
	\draw[color=black] (3.0,0.4) node[right] {\small + HCO};
	
	\draw[color=black] (-3.0,1.4) node[below] {CHD$_2$O};
	\draw[color=black] (0,1.4) node[below] {CH$_2$DO};
	\draw[color=black] (3.0,1.4) node[below] {CH$_3$O};
	
	\draw[color=black] (-1.5,4.4) node[below] {CD$_2$OH~/~CHD$_2$O};
	\draw[color=black] (1.5,3.8) node[below] {CHDOH / CH$_2$DO};
	\draw[color=black] (4.5,4.4) node[below] {CH$_2$OH / CH$_3$O};
	\draw[->, thick] (-1.5,3.8) -- (-1.5,2.6);
	\draw[->, thick] (1.5,3.2) -- (1.5,2.6);
	\draw[->, thick] (4.5,3.8) -- (4.5,2.6);
	\draw[color=red, ->, thick] (0.15,3.2) -- (-0.6,2.7);
	\draw[color=red, ->, thick] (4.1,3.8) -- (2.4,2.7);
	\draw[->, thick] (-1.3,2.6) -- (-1.35, 3.25) -- (-1.25, 3.15) -- (-1.3,3.8);
	\draw[->, thick] (1.7,2.6) -- (1.65, 2.95) -- (1.75, 2.85) -- (1.7,3.2);
	\draw[->, thick] (4.7,2.6) -- (4.65, 3.25) -- (4.75, 3.15) -- (4.7,3.8);
	
	\draw (-1.5,2.6) node[below] {CHD$_2$OH};
	\draw (1.5,2.6) node[below] {CH$_2$DOH};
	\draw (4.5,2.6) node[below] {CH$_3$OH};
	\draw[->, thick] (-2.5,1.4) -- (-2,2.0);
	\draw[->, thick] (0.5,1.4) -- (1.0,2.0);
	\draw[->, thick] (3.5,1.4) -- (4.0,2.0);
	\draw[->, thick] (-1.7,2.0) -- (-1.95, 1.65) -- (-1.95, 1.75) -- (-2.2,1.4);
	\draw[->, thick] (1.3,2.0) -- (1.05, 1.65) -- (1.05, 1.75) -- (0.8,1.4);
	\draw[->, thick] (4.3,2.0) -- (4.05, 1.65) -- (4.05, 1.75) -- (3.8,1.4);

	\draw[->>, thick] (0.5, 2.4) parabola bend (0,2.6) (-0.5, 2.4);
	\draw[->>, thick] (3.6, 2.4) parabola bend (3,2.6) (2.5, 2.4);
	
	\draw[->, thick] (-1.5,5.0) -- (-1.5,4.4);
	\draw[->, thick] (1.5,5.0) -- (1.5,3.8);
	\draw[->, thick] (4.5,5.0) -- (4.5,4.4);
	\draw[color=red, ->, thick] (0.75,5.0) -- (0,4.5);
	\draw[color=red, ->, thick] (3.9,5.0) -- (2.2,3.9);
	
	\draw (-1.5,5.6) node[below] {D$_2$CO};
	\draw (1.5,5.6) node[below] {HDCO};
	\draw (4.5,5.6) node[below] {H$_2$CO};
	\draw[<<->>, thick] (0.7, 5.45) parabola bend (0,5.65) (-0.7, 5.45);
	\draw[<<->>, thick] (3.7, 5.45) parabola bend (3,5.65) (2.3, 5.45);
	\draw[->, thick, dashed] (0.7, 5.25) -- (-0.7, 5.25);
	\draw[->, thick, dashed] (3.7, 5.25) -- (2.3, 5.25);
	\draw[->, thick] (-1.2, 5.6) --(-0.91, 5.8) -- (-0.89, 5.7) --  (-0.6, 5.9);
	\draw[->, thick] (1.8, 5.6) --(2.11, 5.8) -- (2.09, 5.7) --  (2.4, 5.9);
	\draw[->, thick] (1.2, 5.6) --(0.89, 5.8) -- (0.91, 5.7) --  (0.6, 5.9);
	\draw[->, thick] (4.2, 5.6) --(3.89, 5.8) -- (3.91, 5.7) --  (3.6, 5.9);
	
	\draw[color=black] (0,6.2) node[below] {DCO};
	\draw[color=black] (3, 6.2) node[below] {HCO};
	\draw[->, thick] (1.9,6.5) -- (2.3,6.2);
	\draw[color=red, ->, thick] (1.1,6.5) -- (0.675,6.2);
	\draw (1.5, 7.0) node[below] { CO};
\end{tikzpicture}
\caption{\label{fig-scheme} The simplified chemical pathway for the formation of methyl formate, and its isotopologues, on the ice surface of dust grains. Species enclosed in a black box have been observed in the ISM. Normal arrows represent addition reactions, and hydrogenation when there is no radical noted near, dashed lines are for H-D substitution, double headed arrows show abstraction + addition reactions and thunderbolt arrows are for photo-dissociation reactions. D-atom addition reactions are represented in red. It is the compilation of \cite{Garrod-2006}, \cite{Hidaka-2009} and \cite{Taquet-2014} 
studies. }
\end{figure}
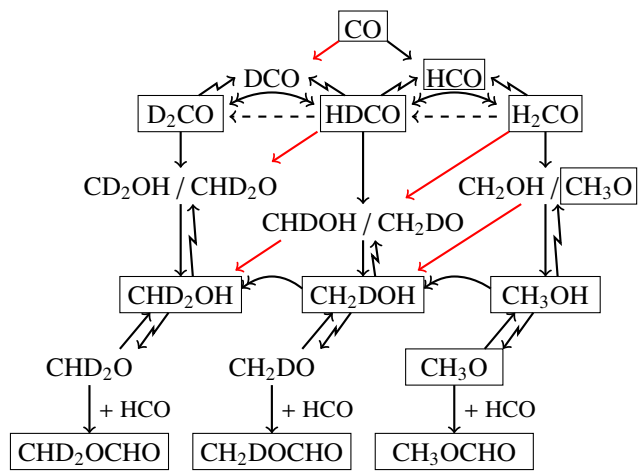

The same chemical network is used during all the physical evolution steps of the system and combines gas phase and grain-surface chemistry. The gas phase network uses the data from the KIDA database, as well as deuterated species, spin states of H$_2$, H$_2^+$ and H$_3^+$ reactions. The grain-surface formalism to treat the chemical processes is based on the multilayer approach developed by \cite{Hasegawa-1993}. It includes the adsorption of gas phases species, the diffusion on the surface, the reaction via the Langmuir--Hinshelwood mechanism, the thermal and chemical desorption, cosmic-ray induced heating and the UV photodissociation and photodesorption. The grain-surface network combines the networks of \cite{Garrod-2006,Garrod-2008,Taquet-2013} and the reaction probability rates of the methanol network are deduced from laboratory experiments \citep{Hidaka-2009}. Figure \ref{fig-scheme} shows the main chemical pathways leading to the formation of methyl formate isotopologues on the grain surface. Methyl formate is assumed in the model to be formed through the radical-radical recombination of HCO and CH$_3$O in warm (20 K $<$ T $<$ 80 K) ices in the protostellar envelope \citep[see][]{Garrod-2006}. Methyl formate deuterated isotopologues are also formed through similar reactions involving HCO, DCO, CH$_3$O, CH$_2$DO, and CHD$_2$O.

 Table 2 shows the abundance ratios and D/H ratios of methyl formate and formaldehyde isotopologues towards IRAS16293A and B, and in the model for the gas located in the inner region of the protostellar envelope (i.e.  $R<$ 50 AU) at two different times in the model labelled "early" and "late" corresponding to $\sim$2$\ \times10^{4}$ yrs after the formation of the central protostar and at the end of the Class 0 stage ($\sim 1\times10^5$ yrs after the protostar formation), respectively. The deuteration enhancement happens during the prestellar phase, when, according to the model, the temperature is the lowest ($T<10$ K) in the inner prestellar core region and increases with the radius as the density and the opacity decrease. At this stage, the temperature in the outer shells of the prestellar envelope is too high and the density too low for CO to freeze out. The presence of CO in the gas inhibits the exothermic gas phase reaction $ \mathrm{H_3^+}+\mathrm{HD}\rightarrow\mathrm{H_2D^+}+\mathrm{HD}$, which is the main deuteration enhancement source at low temperatures. In addition, while the ice is built-up, the deuteration of the freezing species increases with the time. This leads to an increasing D/H ratio profile along the ice layers from the inside to the outside of the ice mantle.
In addition, when the collapse begins and the object turns into a Class 0 protostar, the inner shells of the envelope are richer in deuterated molecules than the outer shells. As the collapse continues, the inner shells of the envelope fall onto the central object first and are replaced by the outer shells. The free-fall collapse assumed in the model induces a fast inward velocity of the envelope shells, limiting their time spent in the hot corino and hence the gas phase chemistry after ice sublimation. Therefore, the D/H ratio predicted in the gas phase of the hot corino decreases with time as it reflects the D/H ratio produced in ices for envelope shells of increasing radii. Towards both components of IRAS16293, the measured D/H ratios are of the same order of magnitude as the "early" time in the model. This suggests that the two protostars are still in their earliest evolutionary (Class 0) stages.

When comparing D/H ratios of singly-deuterated species with their doubly-deuterated forms, the model predicts the D/H ratios to be similar between the two forms (i.e. 7.0\% and 8.9\% at the "early time" and 0.22\% and 0.19\% at the "late time" for the singly- and the doubly-deuterated isotopologues, respectively). However, towards IRAS16293A and B, the D/H ratios of doubly-deuterated methyl formate are higher compared to the singly-deuterated methyl formate by a factor of two to three. The gas phase formation pathways do not contribute significantly to the formation of methyl formate isotopologues in the model and are not able to reach the observed abundances in IRAS16293. This suggests that there are several missing formation pathways on the grain surface in the model.

\cite{Oba-2016, Oba-2016b} found that the H--D substitution reaction in the methyl group of ethanol and dimethyl ether could take place on ice grain surfaces at low temperatures. Similar reactions could also occur on the methyl group of methyl formate, increasing the D/H ratio of CHD$_2$OCHO with respect to CH$_2$DOCHO. In that way, the D--H exchange could be more and more effective as the number of ice layers and the D/H ratio of the gas phase species increase. However, the D/H ratio of CH$_2$DOCHO and CH$_3$OCDO are similar towards both IRAS16293 A and B (see Table 2). \cite{Jorgensen-2018} compared these results to the \cite{Taquet-2014} model and suggested that there was no H--D substitution mechanism on the methyl group, enhancing the D/H ratio of CH$_2$DOCHO with respect to CH$_3$OCDO. 
In this case, the enhanced D/H ratio of CHD$_2$OCHO and CH$_2$DOCHO could be directly inherited from their precursors CHD$_2$OH and CH$_2$DOH, respectively. This is supported by the higher D/H ratio of D$_2$CO in comparison to HDCO, the possible precursors of CHD$_2$OH and CH$_2$DOH, respectively.
Alternatively, if H--D substitutions did occur on ice grain surfaces, increasing significantly the D/H ratio of CH$_2$DOCHO and CHD$_2$OCHO, the H--D substitution could also affect CH$_3$OCDO at a close rate. Further experiments and modelling work are needed to test this.

This new detection shows the importance of determining the D/H ratio of singly- and doubly-deuterated species in order to constrain their formation routes and times. However, the ratios observed in IRAS16293 explore only the chemistry of two sources. In addition, the link between the deuteration level of methanol and methyl formate remains unclear and requires observations of multiply-deuterated isotopologues of these species in a larger sample of protostellar sources, with similar sensitivity and angular resolution provided by interferometric arrays, such as ALMA.

\begin{acknowledgements}
This paper makes use of the following ALMA data: ADS/JAO.ALMA{\#}2013.1.00278.S. ALMA is a partnership of ESO (representing its member states), NSF (USA) and NINS (Japan), together with NRC (Canada) and NSC and ASIAA (Taiwan), in cooperation with the Republic of Chile. The Joint ALMA Observatory is operated by ESO, AUI/NRAO and NAOJ. 
The authors are grateful to Charlotte Vastel for the useful CASSIS software documentation about the synthetic spectrum formalism.
The group of J.~K.~J{\o}rgensen acknowledges support from the European Research Council (ERC) under the European Union’s Horizon 2020 research and innovation programme (grant agreement No 646908) through ERC Consolidator Grant “S4F”. Research at Centre for Star and Planet Formation is funded by the Danish National Research Foundation. 
V.~T. acknowledges the financial support from the European Union’s Horizon 2020 research and innovation programme under the Marie Sklodowska-Curie grant agreement n. 664931.
A.~C. postdoctoral grant is funded by the ERC Starting Grant 3DICE (grant agreement 336474).
M.~N.~D. acknowledges the financial support of the Center for Space and Habitability (CSH) Fellowship and the IAU Gruber Foundation Fellowship.
This research has made use of NASA's Astrophysics Data System as well as community-developed core Python packages for astronomy and scientific computing including 
Astropy (Robitaille et al. 2013), 
Scipy (Jones et al. 2001) and 
Matplotlib (Hunter et al. 2013).
\end{acknowledgements}

\bibliographystyle{aa} 

\bibliography{biblio} 

\appendix

\section{\label{App-spectro}Spectroscopic Data}

The spectroscopic data from the Jet Propulsion Laboratory catalog \citep[JPL;][]{jpl_0} for CH$_3$OCHO, and the Cologne Database for Molecular Spectroscopy \citep[CDMS;][]{cdms_0, cdms_1} entry for CH$_3$O$^{13}$CHO. CH$_3$OCHO data are based on \cite{spec-CH3OCHO_0} with data in the range of the PILS survey from \cite{spec-CH3OCHO_1}, \cite{spec-CH3OCHO_2} and \cite{spec-CH3OCHO_3}. The CH$_3$O$^{13}$CHO entry is based on \cite{spec-CH3O-13-CHO_0}, with measured data in our range from \cite{spec-CH3O-13-CHO_1} and \cite{spec-CH3OCHO_3, spec-CH3O-13-CHO_2}. The spectroscopic data and the partition function for CH$_2$DOCHO are taken from \cite{spec-CH2DOCHO_0}, those for CH$_3$OCDO are taken from \cite{Duan-2015} and those for the doubly-deuterated CHD$_2$OCHO are provided by \cite{spec-CHD2OCHO_0}. The partition function for CH$_3$OCDO is assumed to be the same than CH$_2$DOCHO. The contribution from the excited vibrational levels to the partition function of CH$_3$O$^{13}$CHO and the main isotopologue are determined by \cite{spec-CH3O-13-CHO_cecile}, and are only included in the CH$_3$O$^{13}$CHO database entry. In this study, The vibrational contribution for the partition function of deuterated forms is assumed to be the same as for $^{13}$C-methyl formate. The contribution from the excited vibrational levels is 1.11 and 2.51 for CH$_3$OCHO and 1.31 and 3.85 for both CH$_2$DOCHO and CHD$_2$OCHO, at $T_\mathrm{ex}$ = 115 K and 300 K respectively.

\section{\label{sec-stat-corr}D/H ratio vs abundance ratio}

In the literature, the abundance ratio between the deuterated isotopologue and the main conformer is often presented as the deuterium fractionation. However, when the chemical group containing D isotopes has at least two bonds to H or D atoms, it is not possible to distinguish between the different arrangement of H and D and, hence, to compare the deuteration between different functional groups of the same species (for example between --CH$_3$ and --CHO). 

The probability of having a D atom in a particular site in a chemical group is independent with respect to the other potential sites. Then, the number of complete and undistinguishable chemical groups with $i$ deuterium atoms on $n$ sites is the number of arrangement of the deuterium atoms and the rest in hydrogen atoms. It leads to the relation between the abundance ratio and the D/H ratio:

\begin{equation}
\frac{\mathrm{XH}_{n-i}\mathrm{D}_i}{\mathrm{XH}_n} = \binom{n}{i} \left( \frac{\mathrm{D}}{\mathrm{H}}\right)^i ,
\end{equation}
where $n$ is the number of valence of the X group , $i$ the number of D attached to the X group and $\binom{n}{i} = \frac{n!}{i!(n-i)!}$ the number of arrangement of $i$ into $n$. It leads to:
\begin{equation}
\frac{\mathrm{XH}_{n-i}\mathrm{D}_i}{\mathrm{XH}_{n-j}\mathrm{D}_j} = \frac{\binom{n}{i}}{\binom{n}{j}} \left( \frac{\mathrm{D}}{\mathrm{H}}\right)^{i-j} ,\ i > j\ge 0.
\end{equation}

Table \ref{tab-stat-corr} provides the different relations between abundance ratio and D/H ratio used in this paper. 
It is necessary to have the same number of deuterium in the functional groups in the reference species (i.e. the bottom species in the ratio). For example, the D/H ratio of CHD$_2$OCHO with respect to CH$_3$OCHO and CH$_2$DOCHO gives two different values. Based on the probabilistic definition of the D/H ratio, it is similar to comparing an independent probability of having $A$ with the probability of having $A$ knowing $B$.

\begin{table}[h!]
\caption{\label{tab-stat-corr}Relations between the abundance ratio and the D/H ratio.}
\centering
\begin{tabular}{rcl}
\hline\hline
Abundance ratio & & D/H prescription \\
\hline
HDCO / H$_2$CO &=& 2 (D/H) \\
D$_2$CO / H$_2$CO &=& (D/H)$^2$\\
D$_2$CO / HDCO &=& 1/2 (D/H)$_\mathrm{\ HDCO}$\\
\hline
CH$_2$DOH / CH$_3$OH &=& 3 (D/H)\\
CHD$_2$OH / CH$_3$OH &=& 3 (D/H)$^2$\\
CHD$_2$OH / CH$_2$DOH &=& (D/H)$_\mathrm{\ CH_2DOH}$\\
\hline
CH$_2$DOCHO / CH$_3$OCHO &=& 3 (D/H)\\
CHD$_2$OCHO / CH$_3$OCHO &=& 3 (D/H)$^2$\\
CHD$_2$OCHO / CH$_2$DOCHO &=& (D/H)$_\mathrm{\ CH_2DOCHO}$\\
\hline
\end{tabular}
\end{table}

\onecolumn

\section{\label{App-vciem}Velocity-corrected integrated emission maps}

This section shows the Velocity-corrected INtegrated Emission maps (VINE maps) of the brightest unblended lines of CH$_3$OCHO, CH$_2$DOCHO, CH$_3$OCDO and CHD$_2$OCHO, with three spectra for each source extracted at the continuum peak positions, at a 0\farcs3 offset position and 0\farcs6 offset position (0\farcs5 for the B source).

\begin{figure}[h!]
\centering
\begin{tabular}{p{0.43\textwidth}p{0.25\textwidth}p{0.3\textwidth}}
\centering VINE map &  IRAS16293A &  IRAS16293B
\end{tabular}
\includegraphics[scale=0.55]{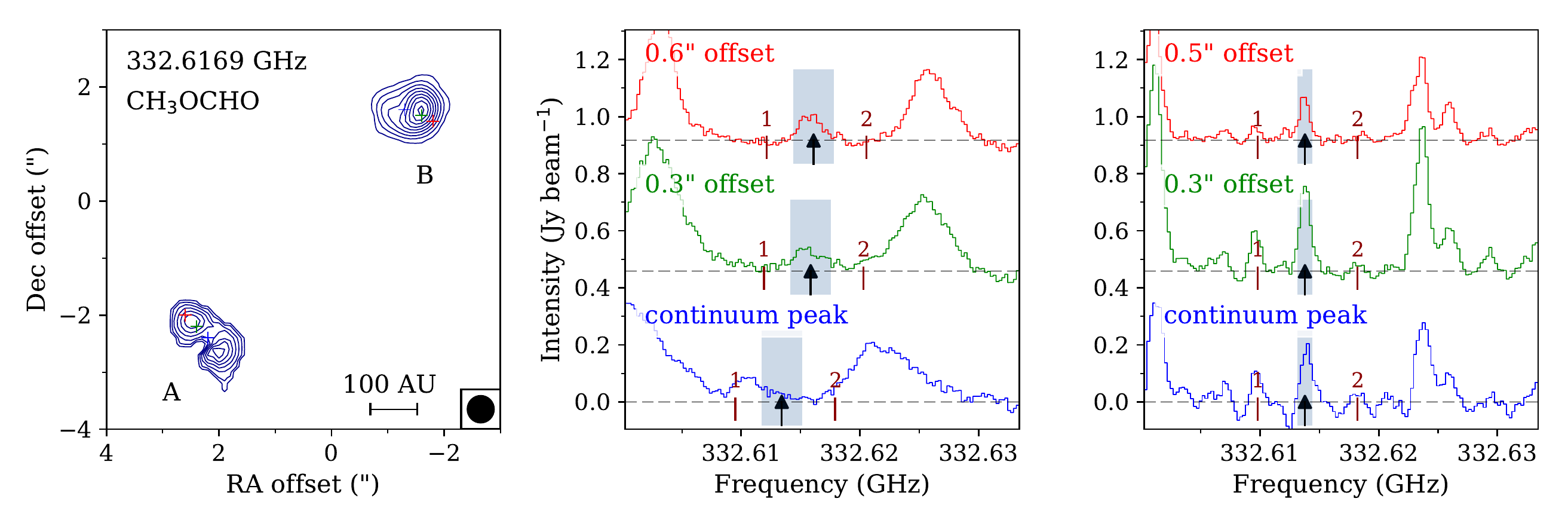}\\
\includegraphics[scale=0.55]{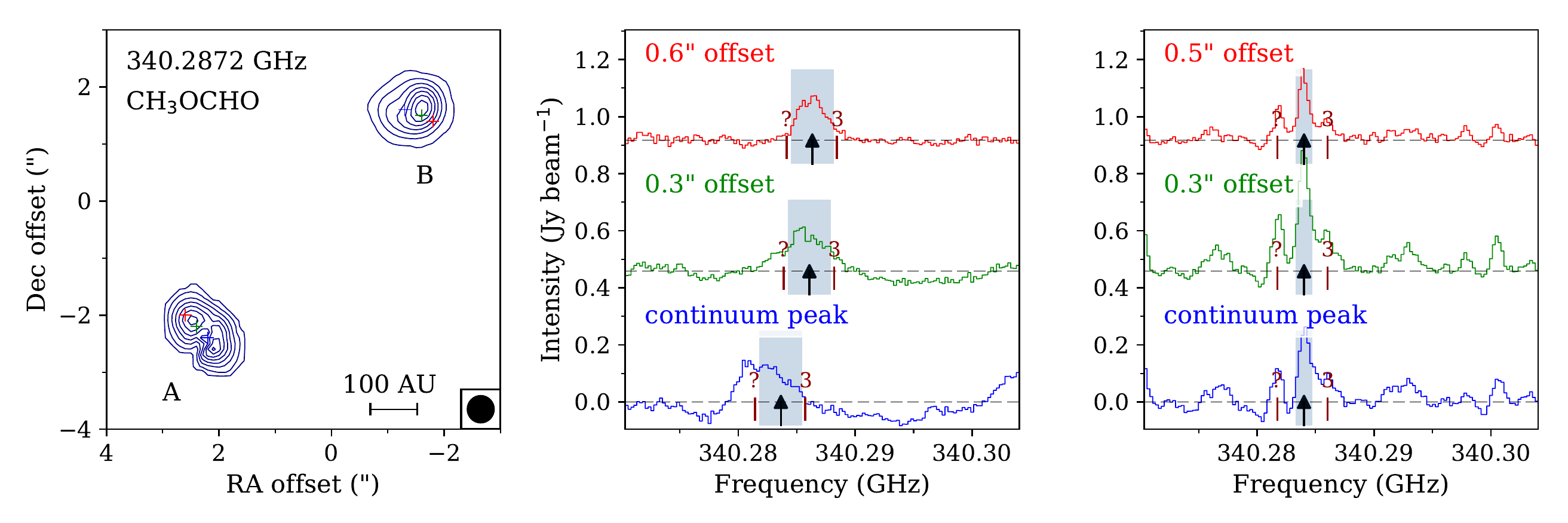}\\
\includegraphics[scale=0.55]{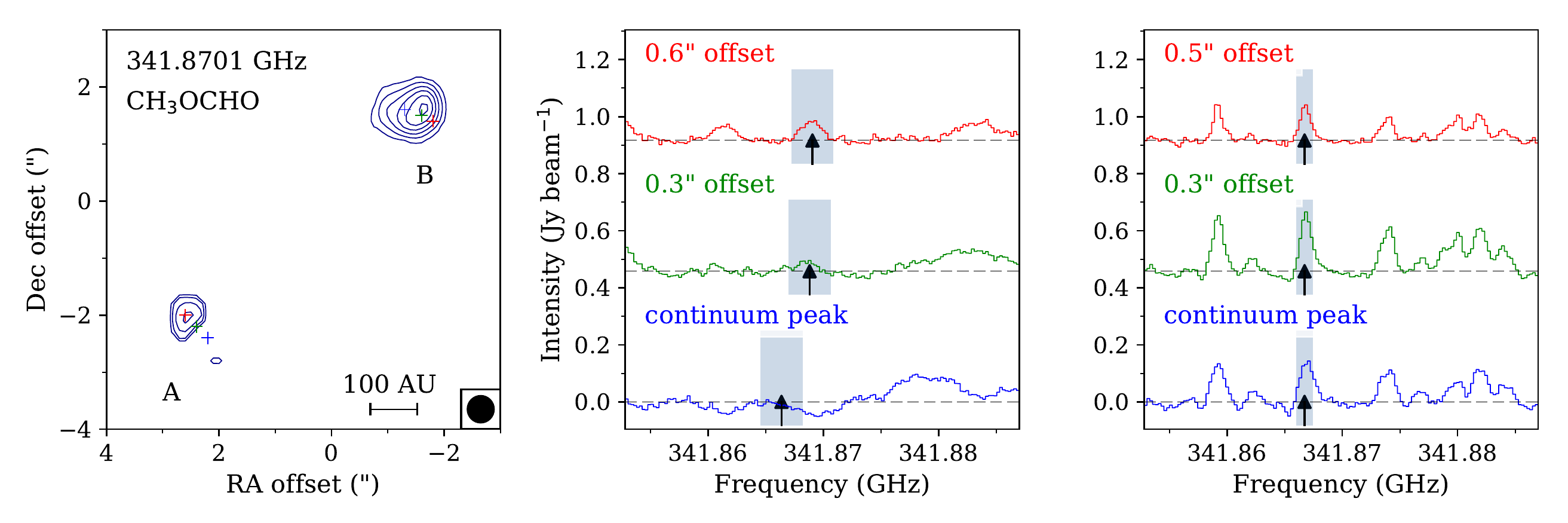}\\
\includegraphics[scale=0.55]{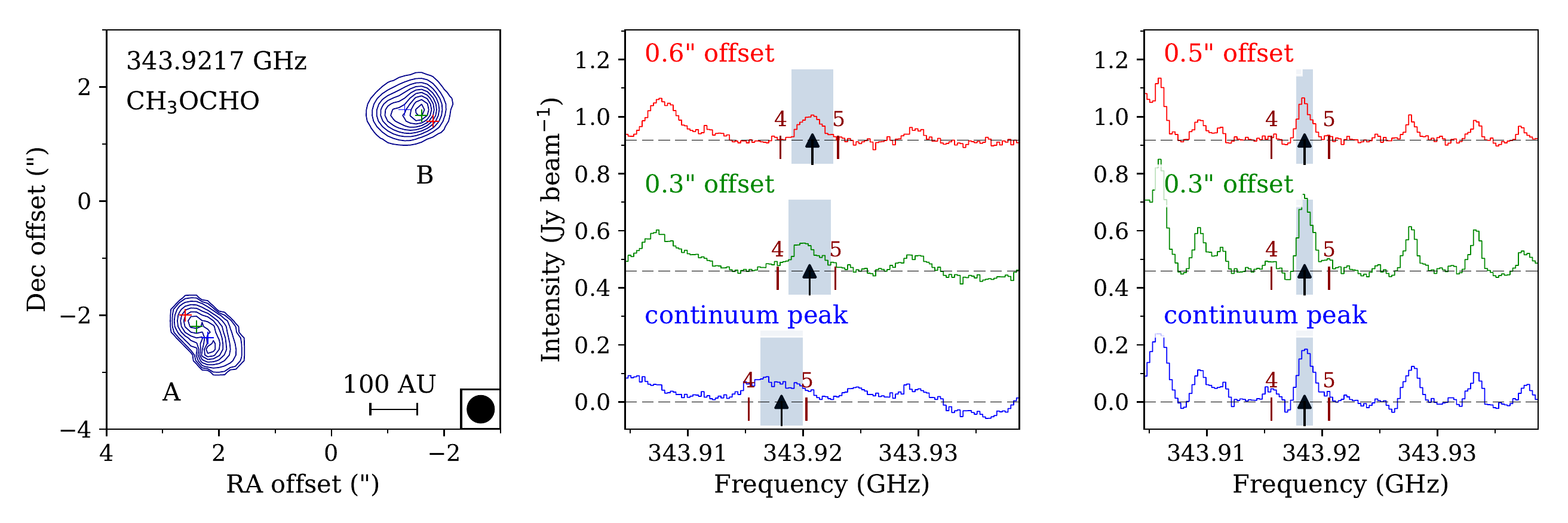}
\caption{VINE maps of the brightest CH$_3$OCHO transitions. Contours levels start at 3$\sigma$ and increase in steps of 3$\sigma$, excepted for the 340,287 GHz line for which the step is 6$\sigma$. The integration range is indicated with a grey rectangle on spectra extracted from three positions, marked with coloured crosses on the map: the continuum peak position in blue, 0\farcs 3 offset position in green and 0\farcs 6 (0\farcs 5 for the B source) offset position in red, towards IRAS16293A (middle) and B (right). The black arrow indicates the rest frequency of the mapped transition. Numbered indicators correspond to other species having a line close to the transition of interest: 1. CH$_2$(OH)CHO $\bullet$ 2. gGg'-(CH$_2$OH)$_2\ \bullet$ 3. CH$_3$CDO $\bullet$ 4. g-n-C$_3$H$_7$CN $\bullet$ 5. S$_2$O (E$_\text{u}$ = 1058~K) $\bullet$ ?. Unidentified.}
\end{figure}

\begin{figure}[h!]
\centering
\begin{tabular}{p{0.43\textwidth}p{0.25\textwidth}p{0.3\textwidth}}
\centering VINE map &  IRAS16293A &  IRAS16293B
\end{tabular}
\includegraphics[scale=0.59]{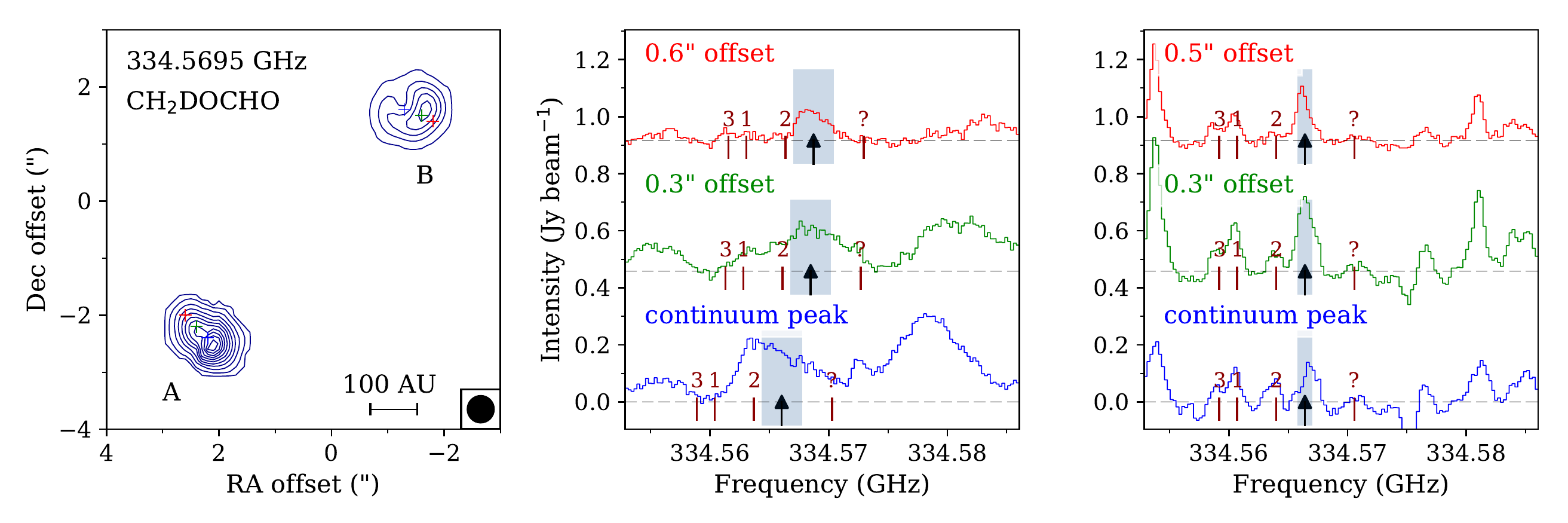}\\
\includegraphics[scale=0.59]{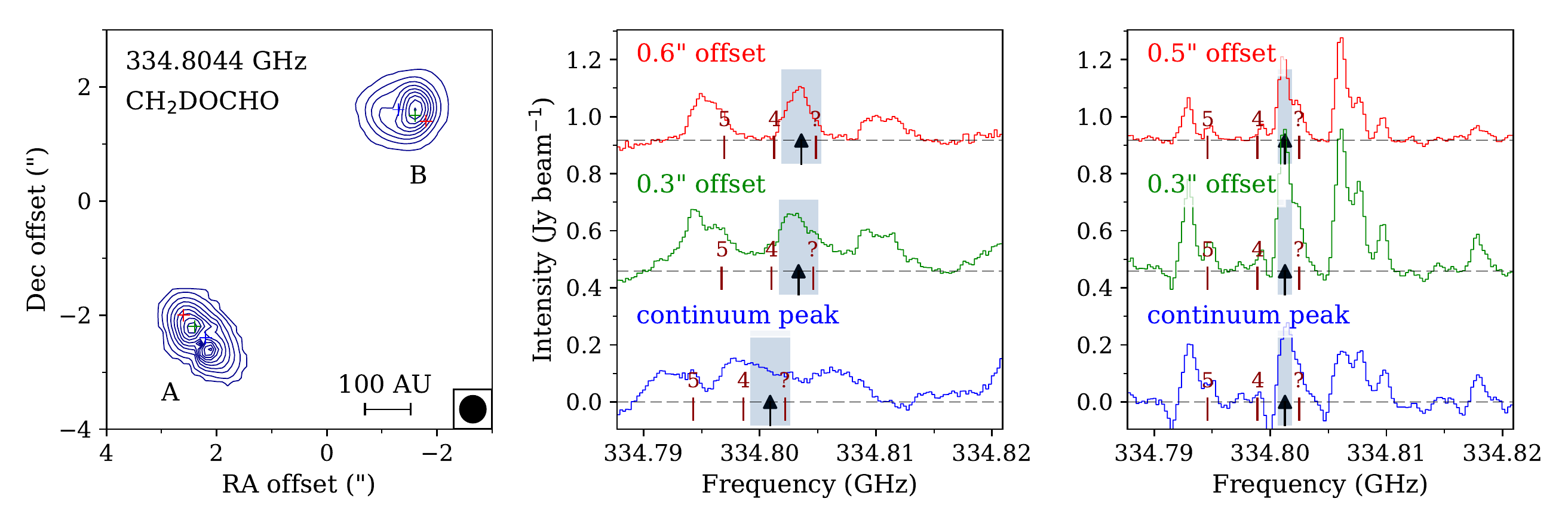}\\
\includegraphics[scale=0.59]{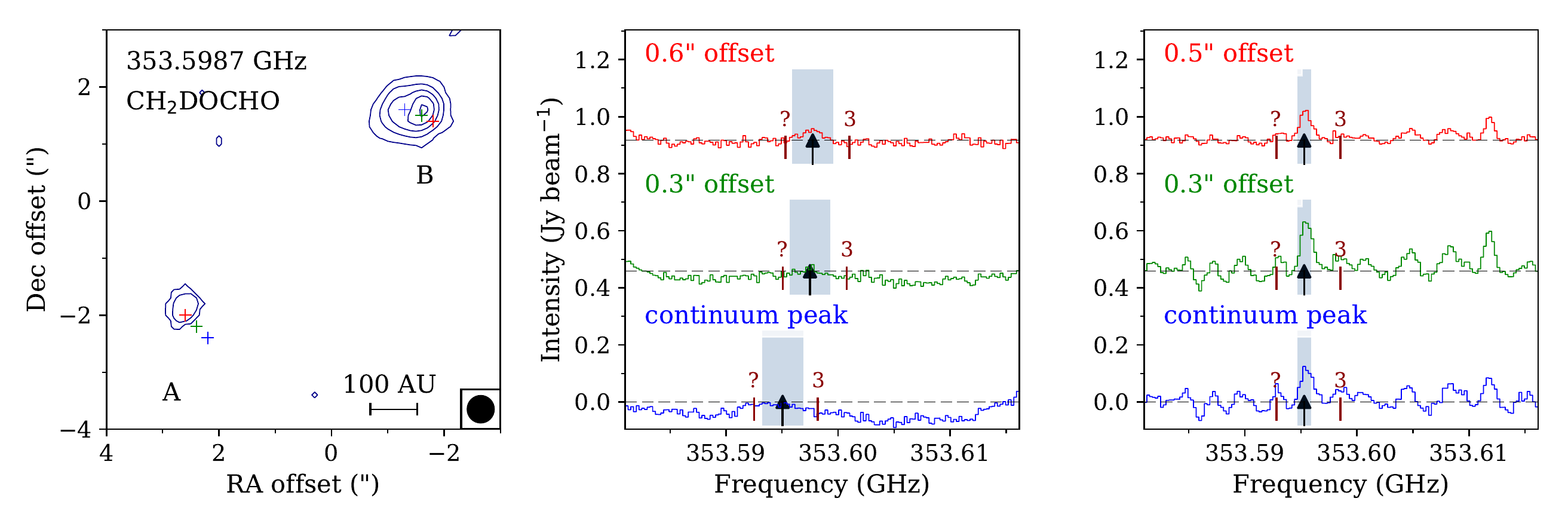}\\
\includegraphics[scale=0.59]{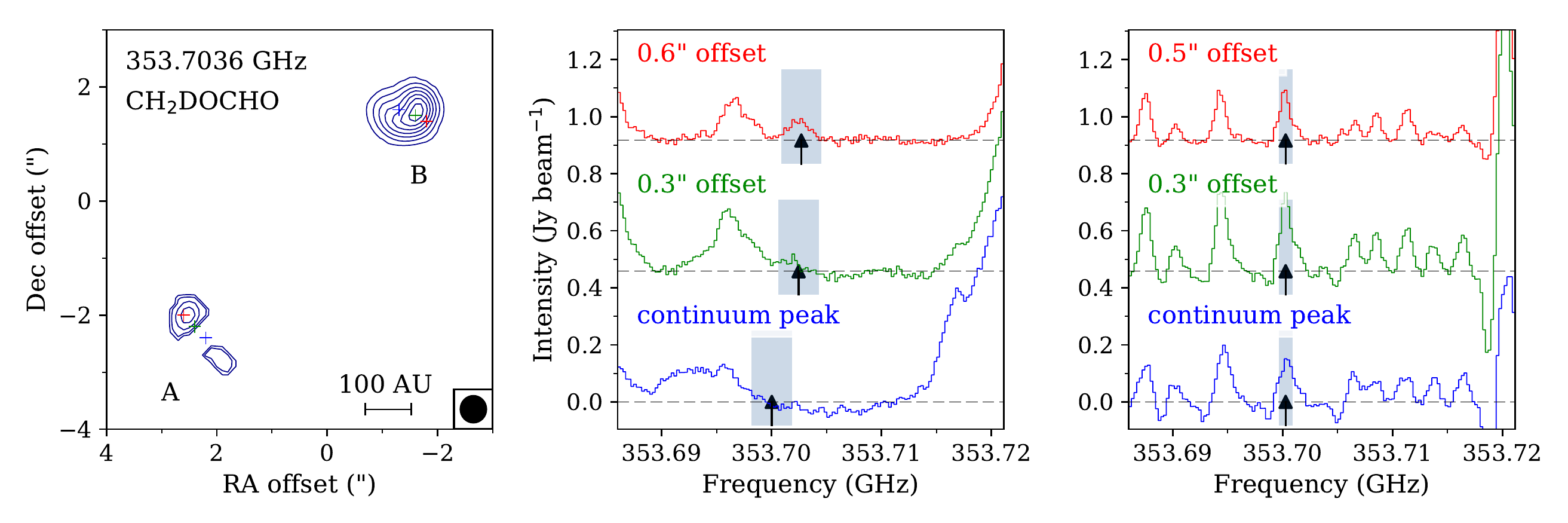}
\caption{VINE maps of the brightest CH$_2$DOCHO transitions. Contours levels start at 3$\sigma$ and increase in steps of 3$\sigma$ for the 353,599 and 353,704 GHz lines and 6$\sigma$ for the 334,569 and 334,804 GHz lines. The integration range is indicated with a grey rectangle on spectra extracted from three positions, marked with coloured crosses on the map: the continuum peak position in blue, 0\farcs 3 offset position in green and 0\farcs 6 (0\farcs 5 for the B source) offset position in red, towards IRAS16293A (middle) and B (right). The black arrow indicates the rest frequency of the mapped transition. Numbered indicators correspond to other species having a line close to the transition of interest: 1. CH$_2$(OH)CHO $\bullet$ 2. CH$_3$CH$_2$OH $\bullet$ 3. CH$_3$O$^{13}$CHO $\bullet$ 4. CH$_3$CHO $\bullet$ 5. CH$_3$COCH$_3\ \bullet$ ?. Unidentified.}
\end{figure}

\begin{figure}[h!]
\centering
\begin{tabular}{p{0.43\textwidth}p{0.25\textwidth}p{0.3\textwidth}}
\centering VINE map &  IRAS16293A &  IRAS16293B
\end{tabular}
\includegraphics[scale=0.59]{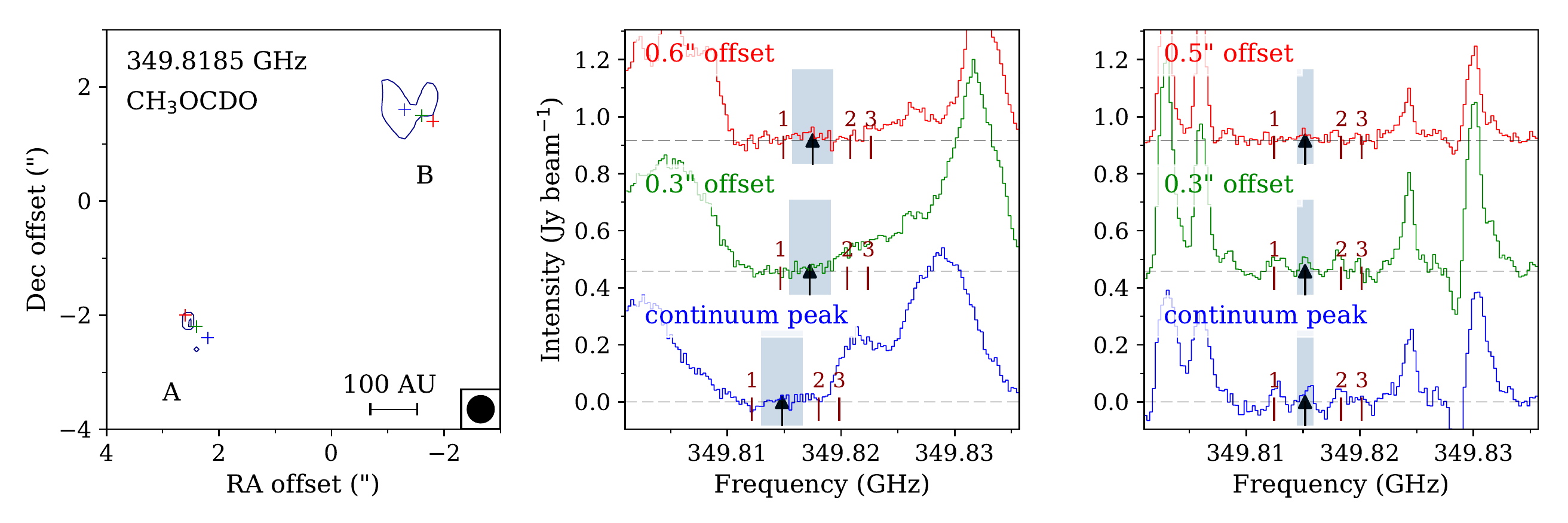}\\
\includegraphics[scale=0.59]{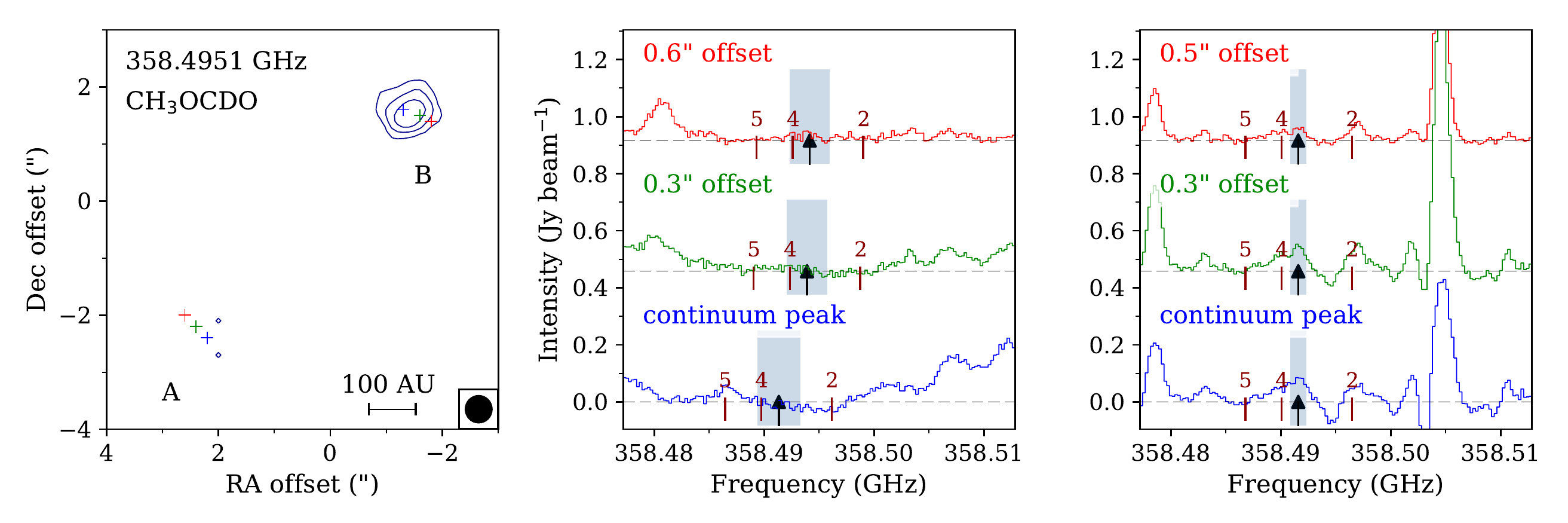}\\
\includegraphics[scale=0.59]{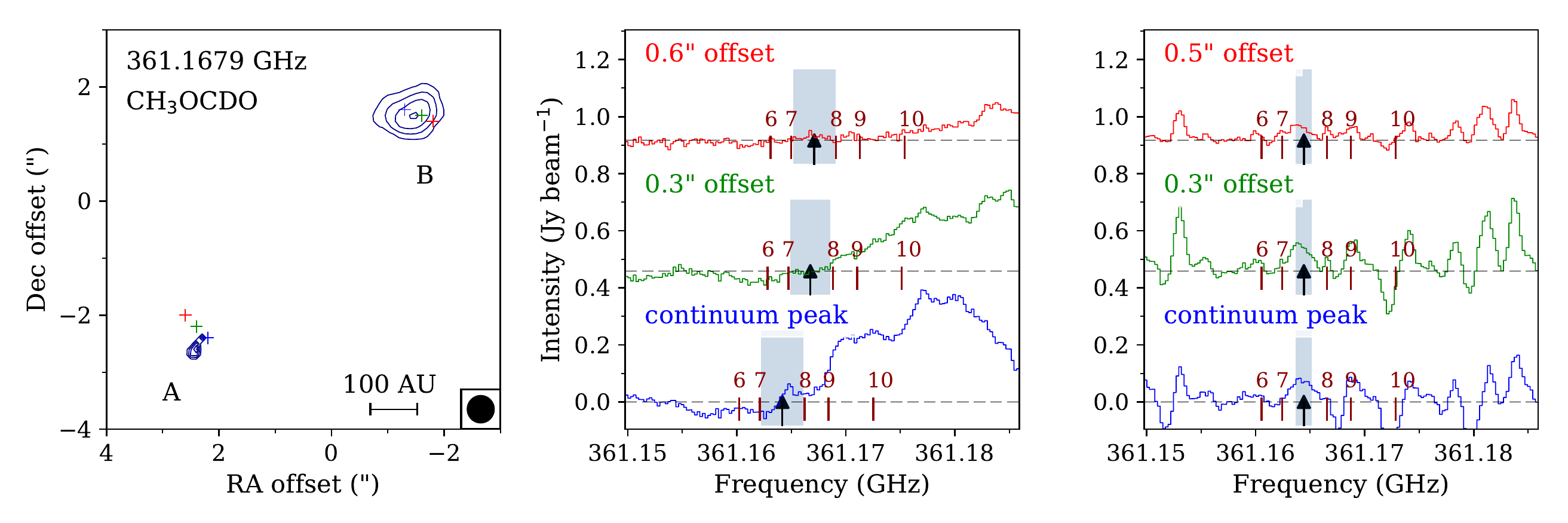}\\
\includegraphics[scale=0.59]{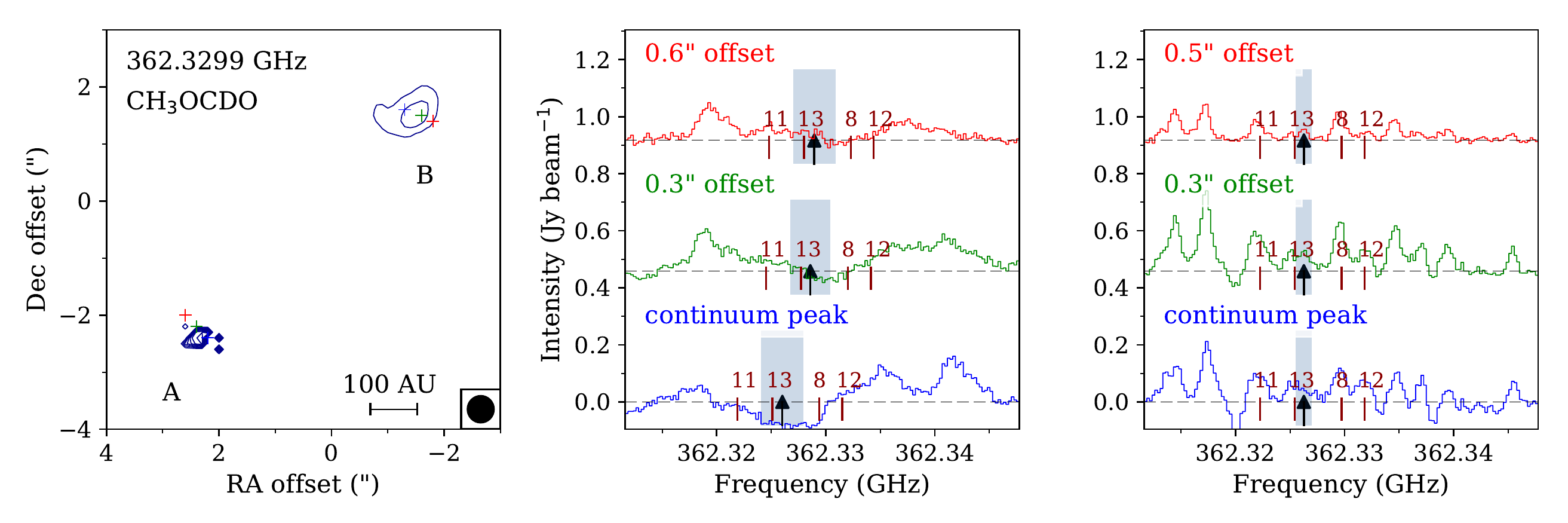}
\caption{VINE maps of the brightest CH$_3$OCDO transitions. Contours levels start at 3$\sigma$ and increase in steps of 3$\sigma$. The integration range is indicated with a grey rectangle on spectra extracted from three positions, marked with coloured crosses on the map: the continuum peak position in blue, 0\farcs 3 offset position in green and 0\farcs 6 (0\farcs 5 for the B source) offset position in red, towards IRAS16293A (middle) and B (right). The black arrow indicates the rest frequency of the mapped transition. Numbered indicators correspond to other species having a line close to the transition of interest: 1. H$^{13}$COOH $\bullet$ 2. gGa-(CH$_2$OH)$_2\ \bullet$ 3. CH$_3$OCH$_3\ \bullet$ 4. CH$_3$CH$_2$OH $\bullet$ 5. HSC $\bullet$ 6. $^{13}$CH$_2$CHCN $\bullet$ 7. CH$_3$CHO $\bullet$ 8. CH$_3$O$^{13}$CHO $\bullet$ 9. CH$_2$(OH)$^{13}$CHO $\bullet$ 10. CH$_2$DOH $\bullet$ 11. CH$_2$(OH)CHO $\bullet$ 12. CH$_3$OCHO $\bullet$ 13. CH$_2$CH$^{13}$CN.}
\end{figure}

\begin{figure}[h!]
\centering
\begin{tabular}{p{0.43\textwidth}p{0.25\textwidth}p{0.3\textwidth}}
\centering VINE map &  IRAS16293A &  IRAS16293B
\end{tabular}
\includegraphics[scale=0.59]{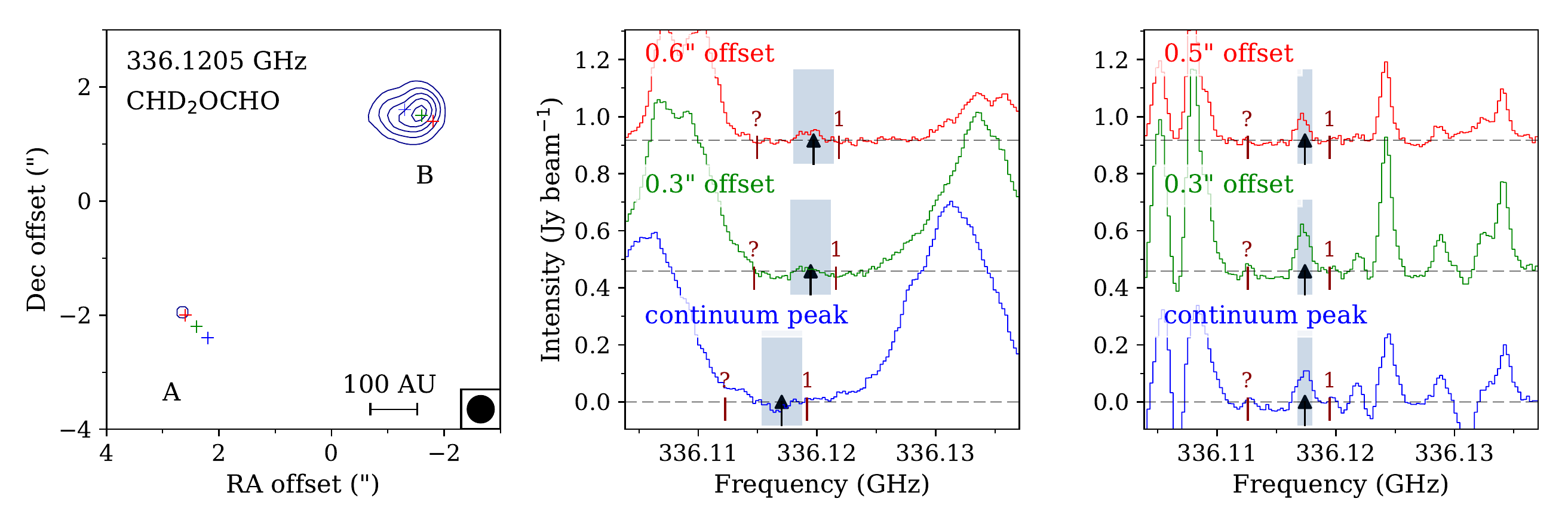}\\
\includegraphics[scale=0.59]{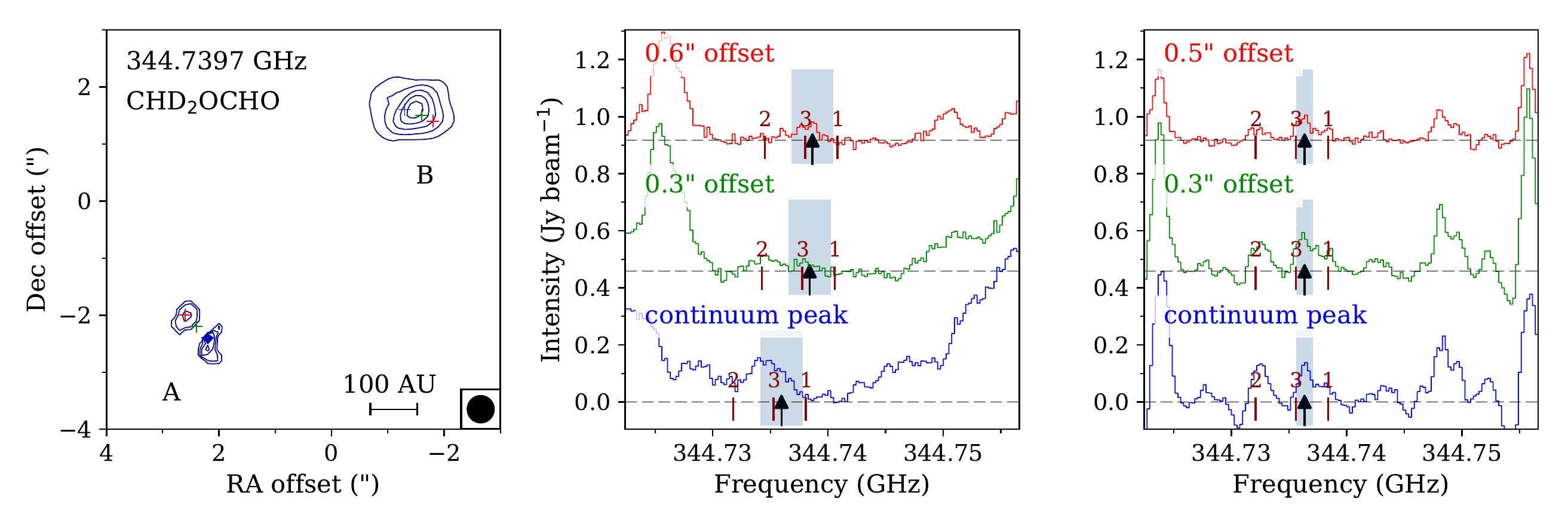}\\
\includegraphics[scale=0.59]{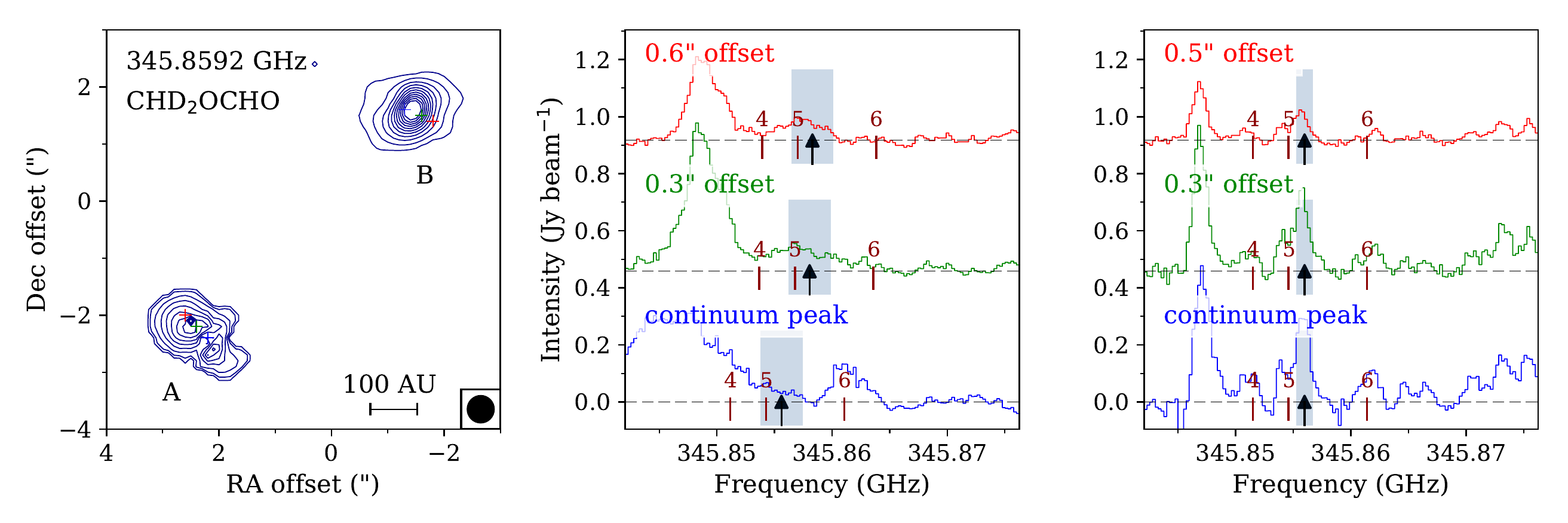}\\
\includegraphics[scale=0.59]{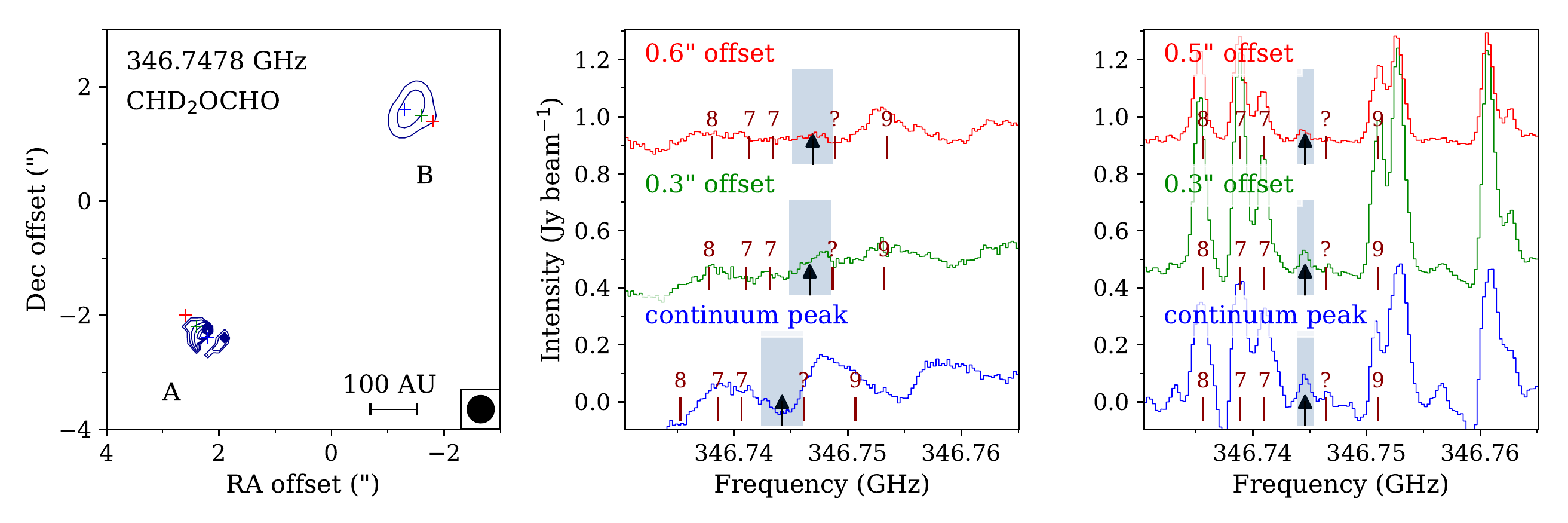}
\caption{VINE maps of the brightest CHD$_2$OCHO transitions. Contours levels start at 3$\sigma$ and increase in steps of 3$\sigma$. The integration range is indicated with a grey rectangle on spectra extracted from three positions, marked with coloured crosses on the map: the continuum peak position in blue, 0\farcs 3 offset position in green and 0\farcs 6 (0\farcs 5 for the B source) offset position in red, towards IRAS16293A (middle) and B (right). The black arrow indicates the rest frequency of the mapped transition. Numbered indicators correspond to other species having a line close to the transition of interest: 1. aGg'-(CH$_2$OH)$_2\ \bullet$ 2. CH$_3$OCHO $\bullet$ 3. CH$_3$O$^{13}$CHO $\bullet$ 4. a-a-CH$_2$DCH$_2$OH $\bullet$ 5. NH$_2$CHO $\bullet$ 6. $^{13}$CH$_3$OH $\bullet$ 7. CH$_3$CHO $\bullet$ 8. HDCO $\bullet$ 9. CH$_3$COCH$_3$ $\bullet$ ?. Unidentified.}
\end{figure}

\section{\label{App-spec}CH$_3$OCHO, CH$_2$DOCHO and CH$_3$OCDO towards IRAS16293A}

This section presents the brightest fitted lines for CH$_3$OCHO, CH$_2$DOCHO and CH$_3$OCDO. 

\begin{figure}[h!]
{CH$_3$OCHO}\\
\includegraphics[scale=0.57]{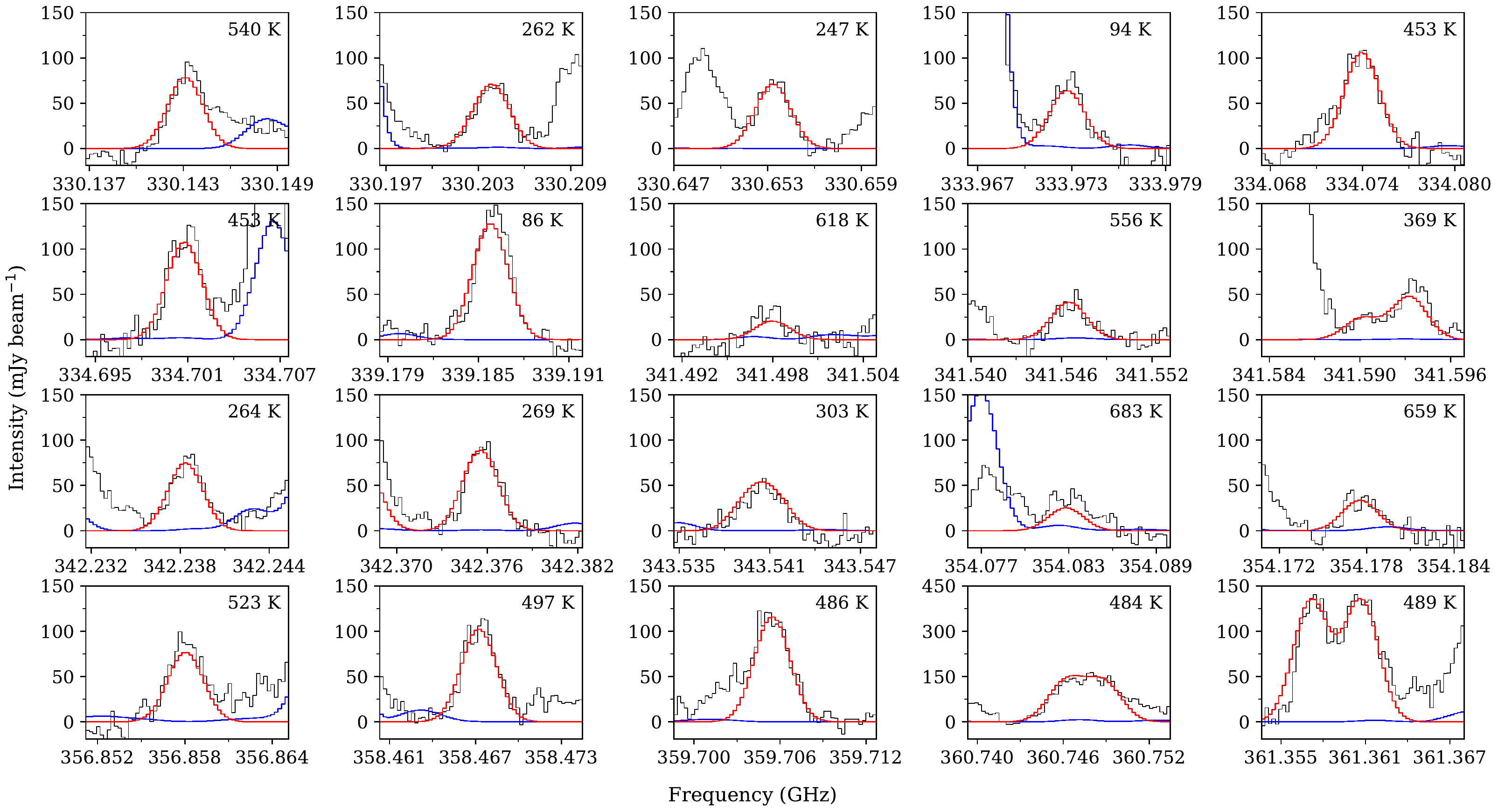}\\
{CH$_2$DOCHO}\\
\includegraphics[scale=0.57]{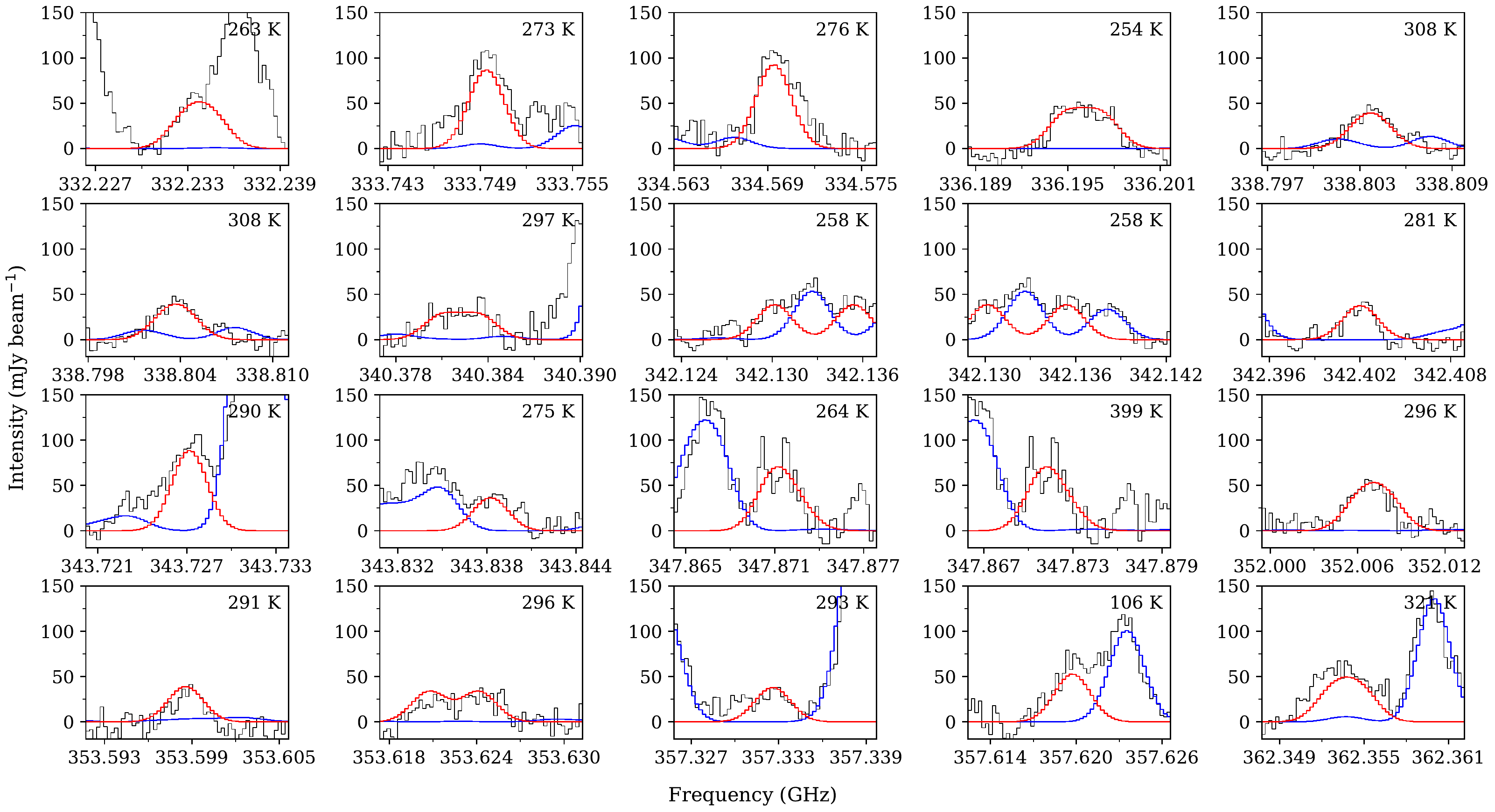}
\caption{20 bright transitions  of CH$_3$OCHO (top) and CH$_2$DOCHO (bottom) towards IRAS16293A. The synthetic spectra are over-plotted in red, a reference spectrum in blue and the data in black. The upper-level energy of each transition is indicated in the top-right corner of each plot.}
\end{figure}

\begin{figure}[h!]
{CH$_3$OCDO}\\
\includegraphics[scale=0.53]{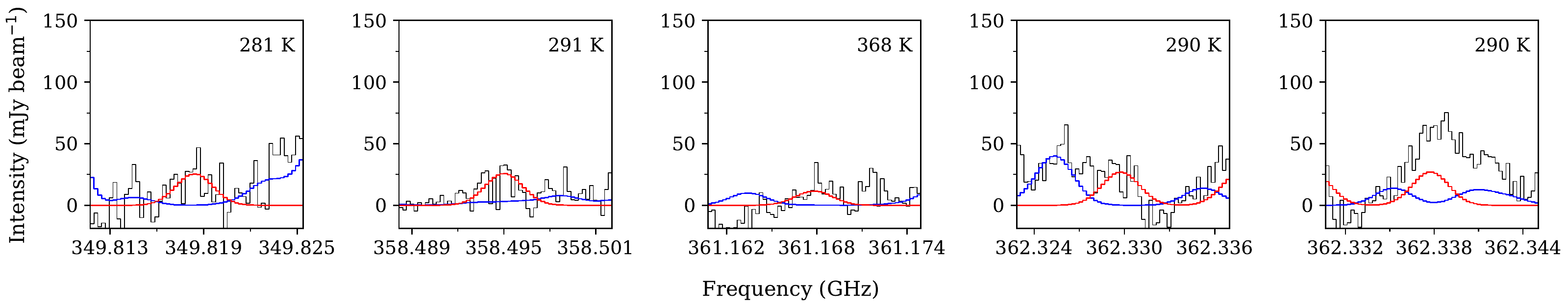}
\caption{The 5 brightest transitions  of CH$_3$OCDO towards IRAS16293A. The synthetic spectra are over-plotted in red, a reference spectrum in blue and the data in black. The upper-level energy of each transition is indicated in the top-right corner of each plot.}
\end{figure}

\section{\label{App-d-spec}Observed CHD$_2$OCHO lines}

This section shows the brightest fitted lines for CHD$_2$OCHO, towards IRAS16293 A and B and the list of plotted transitions. Most of the lines are in fact a combination of unresolved hyperfine transitions, explaining the size of the list compared to the number of lines in the spectra.

\begin{figure}[h!]
{IRAS~16293--2422 A}\\
\includegraphics[scale=0.53]{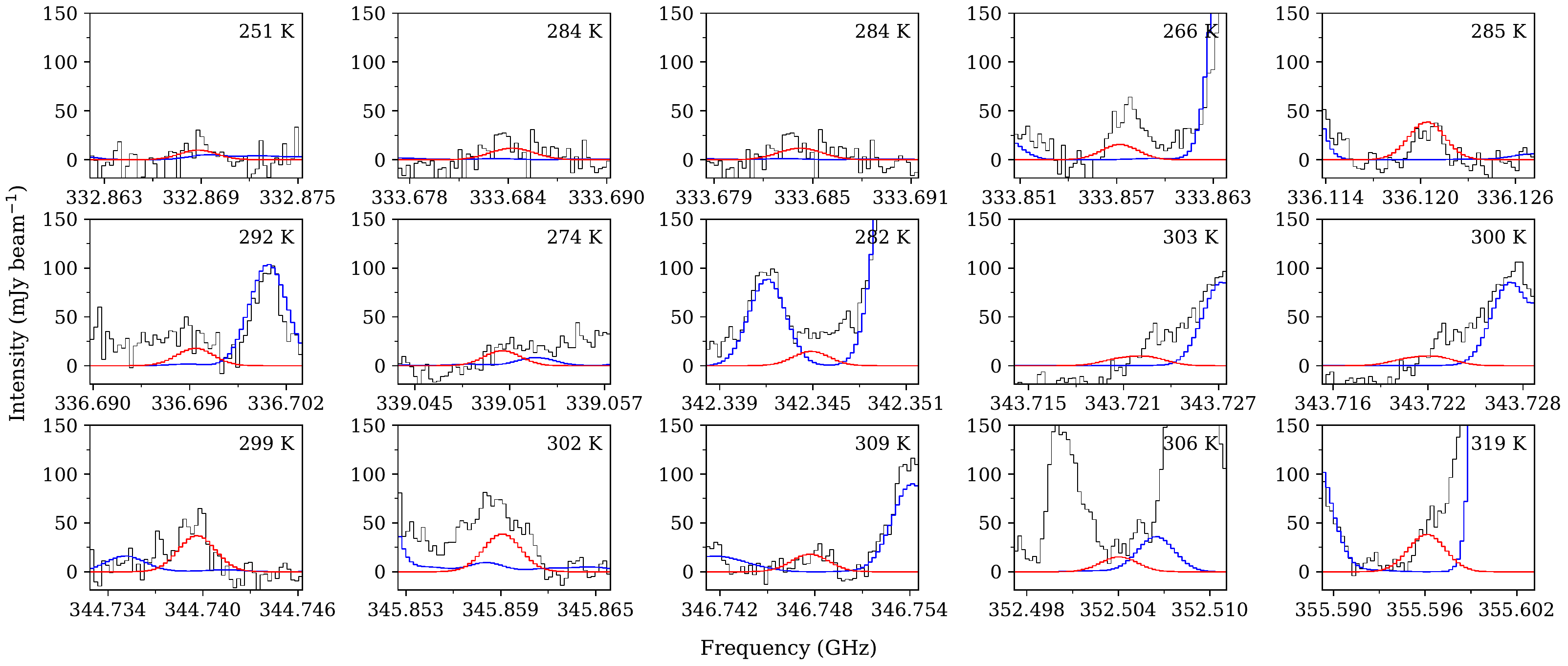}\\
{ IRAS~16293--2422 B}\\
\includegraphics[scale=0.53]{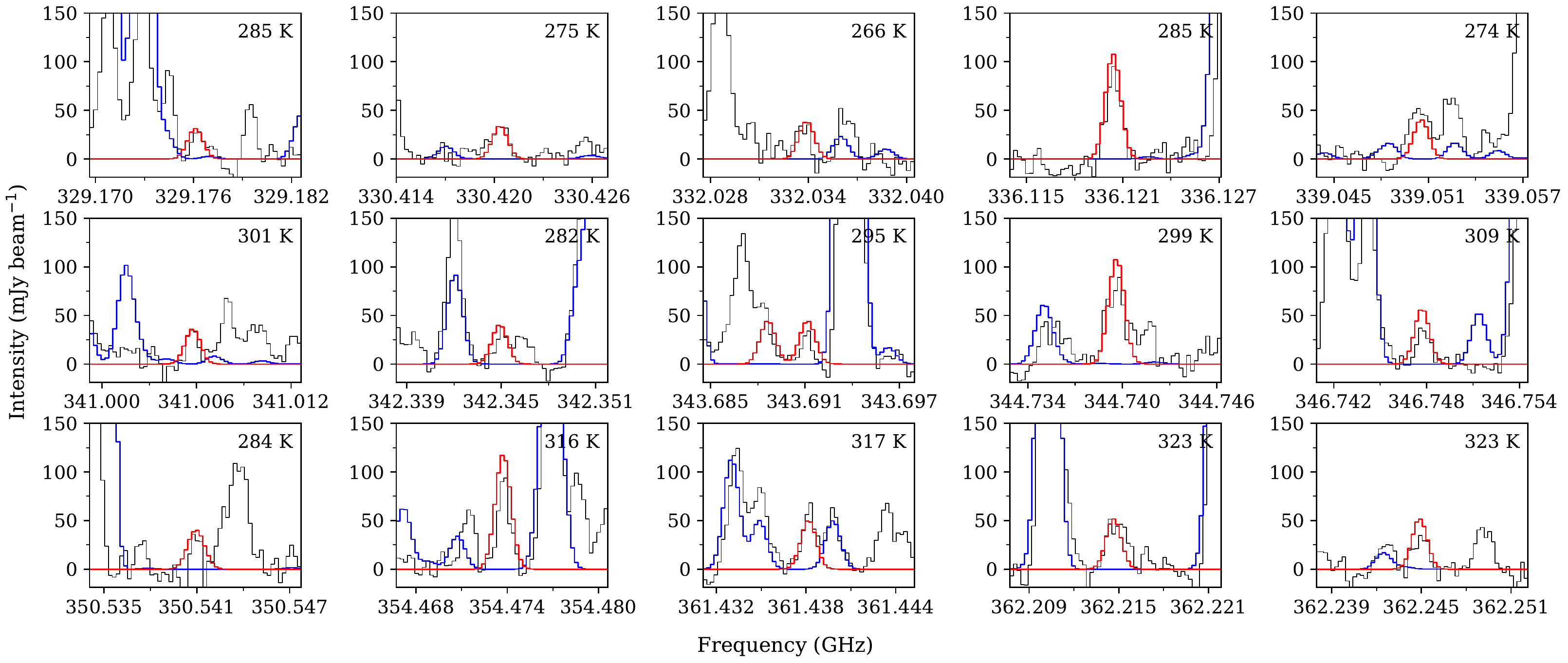}
\caption{\label{App-fig-d-spec}The 15 brightest CHD$_2$OCHO transitions towards IRAS16293A (top) and IRAS16293B (bottom). The synthetic spectra are over-plotted in red, a reference spectrum in blue and the data in black. The upper-level energy of each transition is indicated in the top-right corner of each plot.}
\end{figure}

\subsection{Line list}

Table \ref{App-linelist} shows only the lines used in the minimisation (see section \ref{sec_results}), which corresponds to only a small fraction of the total number of lines in the frequency range of the observations. In total there are 1963 lines for CHD$_2$OCHO and 2137, 2226 and 134 lines for CH$_3$OCHO, CH$_2$DOCHO and CH$_3$OCDO, respectively. 

\begin{longtable}{ccccc}
\caption{\label{App-linelist}Line list of the observed CHD$_2$OCHO transitions used in the fit.}\\
\hline\hline
Transition $J(K_{\mathrm{a}},\ K_{\mathrm{c}})$ -- $J'(K_{\mathrm{a}}',\ K_{\mathrm{c}}')$ & $\mathrm{\nu}$ (GHz) & $g_\mathrm{u}$ & $E_{\mathrm{u}}$ (K) & $A_{\mathrm{ul}}$ (s$^{-1}$) \\
\hline
\endfirsthead
\caption{continued.}\\
\hline\hline
Transition $J(K_{\mathrm{a}},\ K_{\mathrm{c}})\ p$ -- $J'(K_{\mathrm{a}}',\ K_{\mathrm{c}}')\ p'$ & $\mathrm{\nu}$ (GHz) & $g_\mathrm{u}$ & $E_{\mathrm{u}}$ (K) & $A_{\mathrm{ul}}$ (s$^{-1}$) \\
\hline
\endhead
\hline
\endfoot
30(5,26) o$^-$ -- 29(5,25) o$^-\ ^($\footnote{
The CHD$_2$OCHO torsional pattern consists of three non-degenerate sublevels, indicated by the $p$ quantum number. A first 10 cm$^{-1}$ energy symmetry-splitting separates the $i$ and $o$ states, so-called "H-in-plane" and "H-out-plane" respectively. The lower torsional level  $o$ splits again by tunneling into two sublevels, $o^+$ and $o^-$ 10.8 MHz above the first one \citep[for more details, see ][]{spec-CHD2OCHO_0}}$^)$ & 332.0340 & 61 & 266 & $5.305\times10^{-4}$ \\
30(5,26) o$^+$ -- 29(5,25) o$^+$ & 332.0340 & 61 & 266 & $5.305\times10^{-4}$ \\
31(3,28) o$^-$ -- 30(3,27) o$^-$ & 333.1965 & 63 & 273 & $5.405\times10^{-4}$ \\
31(3,28) o$^+$ -- 30(3,27) o$^+$ & 333.1966 & 63 & 273 & $5.404\times10^{-4}$ \\
30(4,26) o$^+$ -- 29(4,25) o$^+$ & 333.8572 & 61 & 266 & $5.397\times10^{-4}$ \\
30(4,26) o$^-$ -- 29(4,25) o$^-$ & 333.8574 & 61 & 266 & $5.397\times10^{-4}$ \\
32(3,30) o$^-$ -- 31(3,29) o$^-$ & 333.9611 & 65 & 278 & $5.490\times10^{-4}$ \\
32(3,30) o$^+$ -- 31(3,29) o$^+$ & 333.9613 & 65 & 278 & $5.490\times10^{-4}$ \\
33(1,32) o$^-$ -- 32(2,31) o$^-$ & 335.0038 & 67 & 282 & $1.071\times10^{-4}$ \\
33(2,32) o$^-$ -- 32(2,31) o$^-$ & 335.0039 & 67 & 282 & $5.594\times10^{-4}$ \\
33(1,32) o$^-$ -- 32(1,31) o$^-$ & 335.0040 & 67 & 282 & $5.594\times10^{-4}$ \\
33(1,32) o$^+$ -- 32(2,31) o$^+$ & 335.0041 & 67 & 282 & $1.071\times10^{-4}$ \\
33(2,32) o$^-$ -- 32(1,31) o$^-$ & 335.0041 & 67 & 282 & $1.071\times10^{-4}$ \\
33(2,32) o$^+$ -- 32(2,31) o$^+$ & 335.0042 & 67 & 282 & $5.594\times10^{-4}$ \\
33(1,32) o$^+$ -- 32(1,31) o$^+$ & 335.0042 & 67 & 282 & $5.594\times10^{-4}$ \\
33(2,32) o$^+$ -- 32(1,31) o$^+$ & 335.0043 & 67 & 282 & $1.071\times10^{-4}$ \\
32(1,31) i -- 31(2,30) i & 335.1566 & 65 & 288 & $1.168\times10^{-4}$ \\
32(2,31) i -- 31(2,30) i & 335.1567 & 65 & 288 & $5.492\times10^{-4}$ \\
32(1,31) i -- 31(1,30) i & 335.1567 & 65 & 288 & $5.492\times10^{-4}$ \\
32(2,31) i -- 31(1,30) i & 335.1567 & 65 & 288 & $1.168\times10^{-4}$ \\
34(0,34) o$^-$ -- 33(1,33) o$^-$ & 336.1205 & 69 & 285 & $1.208\times10^{-4}$ \\
34(1,34) o$^-$ -- 33(1,33) o$^-$ & 336.1205 & 69 & 285 & $5.704\times10^{-4}$ \\
34(0,34) o$^-$ -- 33(0,33) o$^-$ & 336.1205 & 69 & 285 & $5.704\times10^{-4}$ \\
34(1,34) o$^-$ -- 33(0,33) o$^-$ & 336.1205 & 69 & 285 & $1.208\times10^{-4}$ \\
34(0,34) o$^+$ -- 33(1,33) o$^+$ & 336.1205 & 69 & 285 & $1.208\times10^{-4}$ \\
34(1,34) o$^+$ -- 33(1,33) o$^+$ & 336.1205 & 69 & 285 & $5.704\times10^{-4}$ \\
34(0,34) o$^+$ -- 33(0,33) o$^+$ & 336.1205 & 69 & 285 & $5.704\times10^{-4}$ \\
34(1,34) o$^+$ -- 33(0,33) o$^+$ & 336.1205 & 69 & 285 & $1.208\times10^{-4}$ \\
33(0,33) i -- 32(1,32) i & 336.6965 & 67 & 292 & $1.312\times10^{-4}$ \\
33(1,33) i -- 32(1,32) i & 336.6965 & 67 & 292 & $5.629\times10^{-4}$ \\
33(0,33) i -- 32(0,32) i & 336.6965 & 67 & 292 & $5.629\times10^{-4}$ \\
33(1,33) i -- 32(0,32) i & 336.6965 & 67 & 292 & $1.312\times10^{-4}$ \\
30(12,19) o$^+$ -- 29(12,18) o$^+$ & 337.9397 & 61 & 339 & $4.885\times10^{-4}$ \\
30(12,18) o$^+$ -- 29(12,17) o$^+$ & 337.9397 & 61 & 339 & $4.885\times10^{-4}$ \\
30(12,19) o$^-$ -- 29(12,18) o$^-$ & 337.9401 & 61 & 339 & $4.887\times10^{-4}$ \\
30(12,18) o$^-$ -- 29(12,17) o$^-$ & 337.9402 & 61 & 339 & $4.885\times10^{-4}$ \\
30(6,25) o$^-$ -- 29(6,24) o$^-$ & 339.0506 & 61 & 274 & $5.609\times10^{-4}$ \\
30(6,25) o$^+$ -- 29(6,24) o$^+$ & 339.0507 & 61 & 274 & $5.609\times10^{-4}$ \\
31(5,27) o$^+$ -- 30(5,26) o$^+$ & 341.8847 & 63 & 282 & $5.800\times10^{-4}$ \\
31(5,27) o$^-$ -- 30(5,26) o$^-$ & 341.8848 & 63 & 282 & $5.800\times10^{-4}$ \\
32(4,29) o$^-$ -- 31(4,28) o$^-$ & 342.7908 & 65 & 289 & $5.892\times10^{-4}$ \\
32(4,29) o$^+$ -- 31(4,28) o$^+$ & 342.7909 & 65 & 289 & $5.892\times10^{-4}$ \\
29(6,23) o$^-$ -- 28(6,22) o$^-$ & 342.8648 & 59 & 261 & $5.838\times10^{-4}$ \\
29(6,23) o$^+$ -- 28(6,22) o$^+$ & 342.8651 & 59 & 261 & $5.838\times10^{-4}$ \\
32(3,29) o$^-$ -- 31(3,28) o$^-$ & 342.8687 & 65 & 289 & $5.897\times10^{-4}$ \\
32(3,29) o$^+$ -- 31(3,28) o$^+$ & 342.8688 & 65 & 289 & $5.897\times10^{-4}$ \\
33(2,31) o$^-$ -- 32(2,30) o$^-$ & 343.6912 & 67 & 295 & $5.991\times10^{-4}$ \\
33(2,31) o$^+$ -- 32(2,30) o$^+$ & 343.6914 & 67 & 295 & $5.991\times10^{-4}$ \\
32(2,30) i -- 31(3,29) i & 343.7220 & 65 & 300 & $1.123\times10^{-4}$ \\
32(3,30) i -- 31(3,29) i & 343.7228 & 65 & 300 & $5.868\times10^{-4}$ \\
32(2,30) i -- 31(2,29) i & 343.7235 & 65 & 300 & $5.868\times10^{-4}$ \\
32(3,30) i -- 31(2,29) i & 343.7244 & 65 & 300 & $1.123\times10^{-4}$ \\
34(1,33) o$^-$ -- 33(2,32) o$^-$ & 344.7396 & 69 & 299 & $1.173\times10^{-4}$ \\
34(2,33) o$^-$ -- 33(2,32) o$^-$ & 344.7396 & 69 & 299 & $6.101\times10^{-4}$ \\
34(1,33) o$^-$ -- 33(1,32) o$^-$ & 344.7396 & 69 & 299 & $6.101\times10^{-4}$ \\
34(2,33) o$^-$ -- 33(1,32) o$^-$ & 344.7397 & 69 & 299 & $1.173\times10^{-4}$ \\
34(1,33) o$^+$ -- 33(2,32) o$^+$ & 344.7398 & 69 & 299 & $1.173\times10^{-4}$ \\
34(2,33) o$^+$ -- 33(2,32) o$^+$ & 344.7398 & 69 & 299 & $6.101\times10^{-4}$ \\
34(1,33) o$^+$ -- 33(1,32) o$^+$ & 344.7399 & 69 & 299 & $6.101\times10^{-4}$ \\
34(2,33) o$^+$ -- 33(1,32) o$^+$ & 344.7399 & 69 & 299 & $1.173\times10^{-4}$ \\
33(1,32) i -- 32(2,31) i & 345.2040 & 67 & 305 & $1.283\times10^{-4}$ \\
33(2,32) i -- 32(2,31) i & 345.2040 & 67 & 305 & $6.006\times10^{-4}$ \\
33(1,32) i -- 32(1,31) i & 345.2040 & 67 & 305 & $6.006\times10^{-4}$ \\
33(2,32) i -- 32(1,31) i & 345.2041 & 67 & 305 & $1.283\times10^{-4}$ \\
35(0,35) o$^-$ -- 34(1,34) o$^-$ & 345.8592 & 71 & 302 & $1.318\times10^{-4}$ \\
35(1,35) o$^-$ -- 34(1,34) o$^-$ & 345.8592 & 71 & 302 & $6.218\times10^{-4}$ \\
35(0,35) o$^-$ -- 34(0,34) o$^-$ & 345.8592 & 71 & 302 & $6.218\times10^{-4}$ \\
35(1,35) o$^-$ -- 34(0,34) o$^-$ & 345.8592 & 71 & 302 & $1.318\times10^{-4}$ \\
35(0,35) o$^+$ -- 34(1,34) o$^+$ & 345.8592 & 71 & 302 & $1.318\times10^{-4}$ \\
35(1,35) o$^+$ -- 34(1,34) o$^+$ & 345.8592 & 71 & 302 & $6.218\times10^{-4}$ \\
35(0,35) o$^+$ -- 34(0,34) o$^+$ & 345.8592 & 71 & 302 & $6.218\times10^{-4}$ \\
35(1,35) o$^+$ -- 34(0,34) o$^+$ & 345.8592 & 71 & 302 & $1.318\times10^{-4}$ \\
34(0,34) i -- 33(1,33) i & 346.7478 & 69 & 309 & $1.435\times10^{-4}$ \\
34(1,34) i -- 33(1,33) i & 346.7478 & 69 & 309 & $6.151\times10^{-4}$ \\
34(0,34) i -- 33(0,33) i & 346.7478 & 69 & 309 & $6.151\times10^{-4}$ \\
34(1,34) i -- 33(0,33) i & 346.7478 & 69 & 309 & $1.435\times10^{-4}$ \\
31(12,20) o$^+$ -- 30(12,19) o$^+$ & 349.4138 & 63 & 356 & $5.468\times10^{-4}$ \\
31(12,19) o$^+$ -- 30(12,18) o$^+$ & 349.4139 & 63 & 356 & $5.468\times10^{-4}$ \\
31(12,20) o$^-$ -- 30(12,19) o$^-$ & 349.4142 & 63 & 356 & $5.468\times10^{-4}$ \\
31(12,19) o$^-$ -- 30(12,18) o$^-$ & 349.4143 & 63 & 356 & $5.468\times10^{-4}$ \\
32(4,28) o$^+$ -- 31(4,27) o$^+$ & 352.5493 & 65 & 299 & $6.371\times10^{-4}$ \\
32(4,28) o$^-$ -- 31(4,27) o$^-$ & 352.5495 & 65 & 299 & $6.371\times10^{-4}$ \\
34(3,32) o$^-$ -- 33(3,31) o$^-$ & 353.4151 & 69 & 312 & $6.520\times10^{-4}$ \\
34(3,32) o$^+$ -- 33(3,31) o$^+$ & 353.4153 & 69 & 312 & $6.520\times10^{-4}$ \\
35(1,34) o$^-$ -- 34(2,33) o$^-$ & 354.4737 & 71 & 316 & $1.282\times10^{-4}$ \\
35(2,34) o$^-$ -- 34(2,33) o$^-$ & 354.4737 & 71 & 316 & $6.638\times10^{-4}$ \\
35(1,34) o$^-$ -- 34(1,33) o$^-$ & 354.4737 & 71 & 316 & $6.638\times10^{-4}$ \\
35(2,34) o$^-$ -- 34(1,33) o$^-$ & 354.4738 & 71 & 316 & $1.282\times10^{-4}$ \\
35(1,34) o$^+$ -- 34(2,33) o$^+$ & 354.4739 & 71 & 316 & $1.282\times10^{-4}$ \\
35(2,34) o$^+$ -- 34(2,33) o$^+$ & 354.4740 & 71 & 316 & $6.637\times10^{-4}$ \\
35(1,34) o$^+$ -- 34(1,33) o$^+$ & 354.4740 & 71 & 316 & $6.637\times10^{-4}$ \\
35(2,34) o$^+$ -- 34(1,33) o$^+$ & 354.4740 & 71 & 316 & $1.282\times10^{-4}$ \\
30(6,24) o$^-$ -- 29(6,23) o$^-$ & 354.6897 & 61 & 278 & $6.481\times10^{-4}$ \\
30(6,24) o$^+$ -- 29(6,23) o$^+$ & 354.6898 & 61 & 278 & $6.481\times10^{-4}$ \\
34(1,33) i -- 33(2,32) i & 355.2495 & 69 & 322 & $1.405\times10^{-4}$ \\
34(2,33) i -- 33(2,32) i & 355.2495 & 69 & 322 & $6.552\times10^{-4}$ \\
34(1,33) i -- 33(1,32) i & 355.2495 & 69 & 322 & $6.552\times10^{-4}$ \\
34(2,33) i -- 33(1,32) i & 355.2495 & 69 & 322 & $1.405\times10^{-4}$ \\
36(0,36) o$^-$ -- 35(1,35) o$^-$ & 355.5962 & 73 & 319 & $1.435\times10^{-4}$ \\
36(1,36) o$^-$ -- 35(1,35) o$^-$ & 355.5962 & 73 & 319 & $6.761\times10^{-4}$ \\
36(0,36) o$^-$ -- 35(0,35) o$^-$ & 355.5962 & 73 & 319 & $6.761\times10^{-4}$ \\
36(1,36) o$^-$ -- 35(0,35) o$^-$ & 355.5962 & 73 & 319 & $1.435\times10^{-4}$ \\
36(0,36) o$^+$ -- 35(1,35) o$^+$ & 355.5963 & 73 & 319 & $1.436\times10^{-4}$ \\
36(1,36) o$^+$ -- 35(1,35) o$^+$ & 355.5963 & 73 & 319 & $6.761\times10^{-4}$ \\
36(0,36) o$^+$ -- 35(0,35) o$^+$ & 355.5963 & 73 & 319 & $6.761\times10^{-4}$ \\
36(1,36) o$^+$ -- 35(0,35) o$^+$ & 355.5963 & 73 & 319 & $1.436\times10^{-4}$ \\
32(12,21) o$^+$ -- 31(12,20) o$^+$ & 360.9101 & 65 & 373 & $6.096\times10^{-4}$ \\
32(12,20) o$^+$ -- 31(12,19) o$^+$ & 360.9104 & 65 & 373 & $6.096\times10^{-4}$ \\
32(12,21) o$^-$ -- 31(12,20) o$^-$ & 360.9104 & 65 & 373 & $6.096\times10^{-4}$ \\
32(12,20) o$^-$ -- 31(12,19) o$^-$ & 360.9107 & 65 & 373 & $6.096\times10^{-4}$ \\
33(5,29) o$^+$ -- 32(5,28) o$^+$ & 361.4382 & 67 & 317 & $6.872\times10^{-4}$ \\
33(5,29) o$^-$ -- 32(5,28) o$^-$ & 361.4383 & 67 & 317 & $6.872\times10^{-4}$ \\
34(4,31) o$^-$ -- 33(4,30) o$^-$ & 362.2148 & 69 & 323 & $6.968\times10^{-4}$ \\
34(4,31) o$^+$ -- 33(4,30) o$^+$ & 362.2148 & 69 & 323 & $6.968\times10^{-4}$ \\
34(3,31) o$^-$ -- 33(3,30) o$^-$ & 362.2450 & 69 & 323 & $6.971\times10^{-4}$ \\
34(3,31) o$^+$ -- 33(3,30) o$^+$ & 362.2450 & 69 & 323 & $6.971\times10^{-4}$ \\
\hline
\end{longtable}

\end{document}